\documentclass[pre,aps,showpacs,nofootinbib]{revtex4}

\def\Xint#1{\mathchoice
   {\XXint\displaystyle\textstyle{#1}}%
   {\XXint\textstyle\scriptstyle{#1}}%
   {\XXint\scriptstyle\scriptscriptstyle{#1}}%
   {\XXint\scriptscriptstyle\scriptscriptstyle{#1}}%
   \!\int}
\def\XXint#1#2#3{{\setbox0=\hbox{$#1{#2#3}{\int}$}
     \vcenter{\hbox{$#2#3$}}\kern-.5\wd0}}

\def\dashint{\Xint-}
\usepackage{graphicx}
\usepackage{bm}
\usepackage{amssymb}
\usepackage{amsmath}
\usepackage{euscript}

\begin{document}

\title{On-Shell Description of Stationary Flames}\thanks{This paper is an
improved and extended version of the author's previous work
arXiv:physics/0207034 (unpublished).}

\author{Kirill~A.~Kazakov\thanks{E-mail: $Kirill@theor.phys.msu.su$}}

\affiliation{Department of Theoretical Physics, Physics Faculty,
Moscow State University, \\ 119899, Moscow, Russian Federation}

\begin{abstract}
The problem of non-perturbative description of stationary flames
with arbitrary gas expansion is considered. On the basis of the
Thomson circulation theorem an implicit integral of the flow
equations is constructed. With the help of this integral, a simple
explicit expression for the vortex mode of the burnt gas flow near
the flame front is obtained. Furthermore, a dispersion relation
for the potential mode at the flame front is written down, thus
reducing the initial system of bulk equations and jump conditions
for the flow variables to a set of integro-differential equations
for the flame front position and the flow velocity at the front.
The developed approach is applied to the case of thin flames.
Finally, an asymptotic expansion of the derived equations is
carried out in the case $\theta\to 1$ where $\theta$ is the gas
expansion coefficient, and a single equation for the front
position is obtained in the second post-Sivashinsky approximation.
It is demonstrated, in particular, how the well-known problem of
correct normalization of the front velocity is resolved in the new
approach. It is verified also that in the first post-Sivashinsky
approximation, the new equation reduces to the Sivashinsky-Clavin
equation corrected according to Cambray and Joulin. Analytical
solutions of the derived equations are found, and compared with
the results of numerical simulations.
\end{abstract}

\maketitle


\unitlength=1pt

\noindent

\section{Introduction}

The process of flame propagation presents an extremely complicated
mathematical problem. The governing equations include the
nonlinear flow equations for the fuel and the products of
combustion, as well as the transfer equations governing the heat
conduction and species diffusion inside the flame front.
Fortunately, in practice, an inner flame scale determined by the
latter processes is large compared to the flame front thickness,
implying that the flame can be considered as a gasdynamic
discontinuity. The initial problem is reduced thereby to a purely
hydrodynamic problem of determining the propagation of a surface
of discontinuity in an incompressible fluid, the laws of this
propagation given by the usual Navier-Stokes and continuity
equations complemented by the jump conditions at the surface,
expressing the mass and momentum conservation across the flame
front. The asymptotic methods$^{1-3}$ allow one to express these
conditions in the form of a power series with respect to the small
flame front thickness.

Despite this considerable progress, however, a closed theoretical
description of the flame propagation is still lacking. What is
meant by the term ``closed description'' here is the description
of flame dynamics as dynamics of the flow variables {\it on} the
flame front surface. Reduction of the system of bulk equations and
jump conditions, mentioned above, to this ``surface dynamics''
implies solving the flow equations for the fuel and the combustion
products, satisfying given boundary conditions and the jump
conditions at the flame front, and has only been carried out
asymptotically for the case $\theta\to 1,$ where $\theta$ is the
gas expansion coefficient defined as the ratio of the fuel density
and the density of burnt matter$.^{1,4-6}$

Difficulties encountered in trying to obtain a closed description
of flames are conditioned by the following two crucial aspects:

(1) Characterized by the flow velocities which are typically well
below the speed of sound, deflagration represents an essentially
nonlocal process, in the sense that the flame-induced gas flows,
both up- and downstream, strongly affect the flame front structure
itself. A seeding role in this reciprocal process is played by the
Landau-Darrieus (LD) instability of zero thickness flames$.^{7,8}$
A very important factor of non-locality of the flame propagation
is the vorticity production in the flame, which highly complicates
the flow structure downstream. In particular, the local relation
between pressure and velocity fields upstream, expressed by the
Bernoulli equation, no longer holds for the flow variables
downstream.

(2) Deflagration is a highly nonlinear process which requires an
adequate non-perturba-\\tive description of flames with arbitrary
values of the flame front slope. As a result of development of the
LD-instability, exponentially growing perturbations with arbitrary
wavelengths make any initially smooth flame front configuration
corrugated. Characteristic size of the resulting ``cellular''
structure is of the order of the cutoff wavelength $L_c \sim 20
L_{\rm f}$ given by the linear theory of the LD-instability$;^{3}$
$L_{\rm f}$ is the flame front thickness. The exponential growth
of unstable modes is ultimately suppressed by the nonlinear
effects. Since for arbitrary $\theta$ the governing equations do
not contain small parameters, it is clear that the LD-instability
can only be suppressed by the nonlinear effects if the latter are
not small, and therefore so is the flame front slope.

The stabilizing role of the nonlinear effects is best illustrated
in the case of stationary flame propagation. Numerical
experiments$^{9}$ on 2D flames with $\theta = 6-8$ show that even
in very narrow tubes (tube width of the order $L_c$), typical
values for the flame front slope are about $1.5-2.0.$ Nonlinearity
can be considered small only in the case of small gas expansion,
$\theta\to 1,$ where one has the $O(\theta-1)$ estimate for the
slope, so that it is possible to derive an equation$^{4-6}$ for
the flame front position in the framework of the perturbation
expansion in powers of $(\theta - 1).$

This perturbative method gives results in a reasonable agreement
with the experiment only for flames with $\theta \le 3,$
propagating in very narrow tubes (tube width of the order $L_c$),
so that the front slope does not exceed unity. Flames of practical
importance, however, have $\theta$ up to 10, and propagate in
tubes much wider than $L_c.$ As a result of development of the
LD-instability, such flames turn out to be highly curved, which
leads to a noticeable increase of the flame velocity. In
connection with this, a natural question arises whether it is
possible to develop a non-perturbative approach to the flame
dynamics, closed in the sense mentioned above, which would be
applicable to flames with arbitrary gas expansion.

A deeper root of this problem is the following {\it dilemma:} As
was mentioned above, the flame propagation is an essentially
non-local process; on the other hand, this non-locality itself is
determined by the flame front configuration and the structure of
gas flows near the front, so the question is whether an explicit
bulk structure of the flow downstream is necessary in deriving an
equation for the flame front position. In other words, we look for
an approach that would provide a closed description of flames more
directly, without the need to solve the flow equations explicitly.

The purpose of this paper is to develop such approach in the
stationary case.

The paper is organized as follows. The flow equations and related
results needed in our investigation are displayed in
Sec.~\ref{flowequations}. A formal integral of the flow equations
is derived in Sec.~\ref{integral} on the basis of the Thomson
circulation theorem. With the help of this integral, an expression
for the vortex mode of the burnt gas flow near the flame front is
obtained in Sec.~\ref{structure}. To close the system of equations
relating the flow variables on the flame front, potentiality of
the flow upstream, and of the burnt gas flow after extracting the
vortex mode are to be expressed through the values of the fuel
velocity at the front. This is done in Sec.~\ref{dispersion} in
the form of dispersion relations, using which we obtain an
equation relating the fuel velocity at the front and the front
configuration in Sec.~\ref{mainrel}. Complemented by the relation
defining the local burning rate (the evolution equation), the
found equation provides description of stationary flames, which is
closed in the above-mentioned sense. The developed approach is
applied to the particular case of zero-thickness flames in
Sec.~\ref{zero}. Furthermore, it is shown in Sec.~\ref{ltransport}
how effects related to the finite flame front thickness can be
taken into account in the obtained equation. Finally, the case of
weak gas expansion is considered in Sec.~\ref{smallt}, where a
single equation for the flame front position is obtained in the
third and the fourth orders in $(\theta - 1).$ Analytical
solutions of the third- and fourth-order equations are found in
Secs.~\ref{ansol} and \ref{ansol4}. The results obtained are
discussed in Sec.~\ref{discussion}. Appendix~\ref{appa} contains a
consistency check for the expression of the vortex mode, derived
in Sec.~\ref{structure}. Some auxiliary mathematical results used
in the text are summarized in Appendix~\ref{appb}.

\section{Integral representation of flow dynamics}

As was mentioned in the point (1) of Introduction, an important
factor of the flow non-locality downstream is the vorticity
production in curved flames, which highly complicates relations
between the flow variables. In the presence of vorticity, pressure
is expressed through the velocity field by an integral relation,
its kernel being the Green function of the Laplace operator. It
should be noted, however, that the jump condition for the pressure
across the flame front only serves as the boundary condition for
determining an appropriate Green function, being useless in other
respects. Thus, it is convenient to exclude pressure from our
consideration from the very beginning. The basis for this is
provided by the well-known Thomson circulation theorem. Thus, we
begin in Sec.~\ref{flowequations} with the standard formulation of
the problem of flame propagation, and then construct a formal
implicit solution of the flow equations with the help of this
theorem in Sec.~\ref{integral}.

\subsection{Flow equations}\label{flowequations}

Let us consider a 2D stationary flame propagating in an initially
uniform premixed ideal fluid. Let the Cartesian coordinates
$(x,y)$ be chosen so that the $y$-axis is in the direction of
flame propagation, $y = - \infty$ being in the fresh fuel. It will
be convenient to introduce the following dimensionless variables
$$(\eta,\xi )=(x/R,y/R)\,, \  (w,u) =({\rm v}_{x}/U_{\rm f}\,,
{\rm v}_{y}/U_{\rm f})\,,$$
$$\Pi =(P-{P}_{\rm f}) /{\rho }_{-}{U_{\rm f}}^{2},$$
where $U_{\rm f}$ is the velocity of a plane flame front, $P_{\rm
f}$ is the initial pressure in the fuel far ahead of the flame
front, and $R$ is some characteristic length of the problem (e.g.,
the tube width). The fluid density will be normalized on the fuel
density $\rho_{-}.$

As numerical experiments show, stationary flames exist only in
sufficiently narrow tubes. Hence, assuming the tube walls ideal,
and denoting its width by $b,$ we will deal below with spatially
$2b$-periodic flames. More precisely, given a flame configuration
described by the functions $f(\eta), w(\eta,\xi), u(\eta,\xi),$
$\eta \in [0,+b],$ where $f(\eta)$ denotes the flame front
position, using the boundary conditions $f'=0, w=0$ for $\eta =
0,b,$ we continue this solution to the domain $\eta \in [-b,0]$
according to
\begin{eqnarray}\label{antisym}
f(\eta) = f(-\eta)\,, \qquad w(\eta,\xi) = - w(-\eta,\xi)\,,
\qquad u(\eta,\xi) = u(-\eta,\xi)\,,
\end{eqnarray}
\noindent and then periodically continue it to the whole
$\eta$-axis.

In connection with this procedure the following circumstance
should be emphasized. All subsequent analysis is carried out under
assumption that there exists a short wavelength cut-off for the
flame perturbations. In other words, we consider flames with small
but non-zero front thickness$.^{10}$ Existence of such a cut-off
ensures smoothness of the functions under consideration. In
particular, it prevents development of singularities such as the
edge points occurring at the front of a zero-thickness flame,
which lead to discontinuities in the values of the flow variables
or their derivatives. The end points of the flame front, however,
still represent a potential source of such discontinuities even in
the case of flames with non-zero thickness because of the
possibility of stream line refraction at these points, resulting
in a formation of stagnation zones in the flow of burnt matter
(see Ref.~11 for more detail). Having imposed the boundary
condition $f'(0) = f'(b) = 0$ we thereby exclude this possibility.
In view of what has just been said, it is natural to assume
further that considered as functions of the complex argument,
$f(\eta)$ together with the ``on-shell'' values of the flow
velocity, $w(\eta,f(\eta)),$ $u(\eta,f(\eta)),$ are analytical
functions of $\eta$ in a vicinity of the real axis. This
assumption, simplifying subsequent analysis, is only technical,
and can be weakened if necessary.

As always, we assume that the process of flame propagation is
nearly isobaric. Then the velocity and pressure fields obey the
following equations in the bulk
\begin{eqnarray}\label{flow1}
\frac{\partial v_i}{\partial\zeta_i} &=& 0\,,
\\ v_i\frac{\partial v_k}{\partial\zeta_i}
&=& - \frac{1}{\rho}\frac{\partial\Pi}{\partial\zeta_k}\,, \quad k
= 1,2, \label{flow2}
\end{eqnarray}
\noindent where $ (\zeta_1,\zeta_2) = (\eta,\xi),\, (v_1,v_2) =
(w,u),$ and summation over repeated indices is implied.

Acting on Eq.~(\ref{flow2}) by the operator
$\varepsilon_{kl}\partial/\partial\zeta_l,$ where
$\varepsilon_{ik} = - \varepsilon_{ki},\  \varepsilon_{12} = + 1,$
and using Eq.~(\ref{flow1}), one obtains a 2D version of the
Thomson circulation theorem
\begin{eqnarray}\label{thomson}
v_i\frac{\partial\sigma}{\partial\zeta_i} = 0\,,
\end{eqnarray}
\noindent where
$$\sigma \equiv \frac{\partial u}{\partial\eta}
- \frac{\partial w}{\partial\xi}\,.$$ According to
Eq.~(\ref{thomson}), the vorticity $\sigma$ is conserved along the
stream lines. As a simple consequence of this theorem, one can
find the general solution of the flow equations upstream. Namely,
since the flow is potential at $\xi = - \infty$ ($u = V = {\rm
const},$ $w = 0,$ where $V$ is the velocity of the flame in the
rest frame of reference of the fuel), it is potential for every
$\xi<f(\eta).$ Therefore,
\begin{eqnarray}\label{solup1}
u &=& \sum\limits_{n= - \infty}^{+ \infty} ~u^{(n)}
\exp\left\{\frac{\pi}{b}(|n|\xi + i n\eta)\right\}\,, \quad
u^{(0)} = V\,,
\\ \label{solup2} w &=& \hat{H}(u - u^{(0)})\,,
\end{eqnarray}
\noindent where the linear Hilbert operator $\hat{H}$ is defined
by
\begin{eqnarray}\label{hilbert}
\hat{H}\exp(i k\eta) = i\chi(k)\exp(i k\eta)\,, \quad k\ne 0\,,
\end{eqnarray}
and $$\chi(x) = \left\{
\begin{array}{cc}
+1,& x>0\,,\\
-1,&  x<0\,.
\end{array}
\right.
$$
In the coordinate representation, $\hat{H}$ acts on a function
$a(\eta),$ such that $a(\eta)\to 0 $ for $\eta \to \pm \infty,$
according to $$\left(\hat{H}a\right)(\eta) =
\frac{1}{\pi}~\dashint\limits_{-\infty}^{+\infty}d\tilde{\eta}
\frac{a(\tilde{\eta})}{\tilde{\eta}-\eta}\,.$$

It will be shown in the next section how the Thomson theorem can
be used to obtain a formal integral of the flow equations
downstream.

\subsection{Integration of the flow equations}\label{integral}

Consider the quantity $$a_i(\zeta) = \triangle^{-1} v_i =
\int\limits_{\Sigma}d s~v_i\frac{\ln r}{2\pi} \,,$$ where $r$ is
the distance from an infinitesimal fluid element $d s$ to the
point of observation $\zeta,$ $r^2 = (\zeta_i -
\tilde{\zeta_i})^2,$ and integration is carried over $\Sigma =
\{\tilde{\eta},\tilde{\xi}: \tilde{\xi} > f(\tilde{\eta})\}.$
Taking into account Eq.~(\ref{flow1}), one has for the divergence
of $a_i:$
\begin{eqnarray}\label{diva}
\frac{\partial a_i}{\partial\zeta_i} = \int\limits_{\Sigma}ds~v_i
\partial_i \left(\frac{\ln r}{2\pi}\right)  = -
\int\limits_{\Sigma}ds~v_i\tilde{\partial}_i \left(\frac{\ln
r}{2\pi}\right) = - \int\limits_{\Sigma}ds~\tilde{\partial}_i
\left(v_i\frac{\ln r}{2\pi} \right) = - \int\limits_{\Lambda}d
l_i~v_i\frac{\ln r}{2\pi} \,,
\end{eqnarray}
\noindent where $\tilde{\partial_i} \equiv
\partial/\partial\tilde{\zeta}_i,$ $\Lambda$ boundary of $\Sigma,$
and $d l_i $ its element.

Next, let us calculate $\varepsilon_{ik}\partial_k
\varepsilon_{lm}\partial_l a_m.$ Using Eq.~(\ref{diva}), we find
\begin{eqnarray}
\varepsilon_{1k}\partial_k \varepsilon_{lm}\partial_l a_m &=&
\frac{\partial}{\partial\xi}\left(\frac{\partial
a_2}{\partial\eta} - \frac{\partial a_1}{\partial\xi}\right) =
\frac{\partial}{\partial\eta}\left(- \frac{\partial
a_1}{\partial\eta} - \int\limits_{\Lambda}d l_i~v_i\frac{\ln
r}{2\pi} \right) - \frac{\partial^2 a_1}{\partial\xi^2}
\nonumber\\
&=& - \triangle a_1 - \frac{\partial}{\partial\eta}
\int\limits_{\Lambda}d l_i~v_i\frac{\ln r}{2\pi} \,. \nonumber
\end{eqnarray}
\noindent Analogously,
\begin{eqnarray}
\varepsilon_{2k}\partial_k \varepsilon_{lm}\partial_l a_m = -
\triangle a_2 - \frac{\partial}{\partial\xi}\int\limits_{\Lambda}
d l_i~v_i\frac{\ln r}{2\pi} \,. \nonumber
\end{eqnarray}
\noindent Together, these two equations can be written as
\begin{eqnarray}
\varepsilon_{ik}\partial_k \varepsilon_{lm}\partial_l a_m = -
\triangle a_i - \partial_i\int\limits_{\Lambda} d l_k~v_k\frac{\ln
r}{2\pi} \,. \nonumber
\end{eqnarray}
\noindent Substituting the definition of $a_i$ into the latter
equation, and integrating by parts gives
\begin{eqnarray}\label{vint}
v_i &=& - \varepsilon_{ik}\partial_k \varepsilon_{lm}
\partial_l \triangle^{-1} v_m
- \partial_i\int\limits_{\Lambda}d l_k~v_k\frac{\ln r}{2\pi}
\nonumber\\
&=& \varepsilon_{ik}\partial_k \int\limits_{\Lambda}d
l_l~\varepsilon_{lm} v_m\frac{\ln r}{2\pi} -
\partial_i\int\limits_{\Lambda}d l_k~v_k\frac{\ln r}{2\pi} -
\varepsilon_{ik}\partial_k \int\limits_{\Sigma}d s~\frac{\ln
r}{2\pi} \sigma \,.
\end{eqnarray}
\noindent The first two terms on the right of Eq.~(\ref{vint})
represent the potential component of the fluid velocity$,^{12}$
while the third corresponds to the vortex component. Our aim below
will be to transform the latter to an integral over the flame
front surface. To this end, we decompose $\Sigma$ into elementary
$d s$ as follows.

\begin{figure}
\scalebox{0.8}{\includegraphics{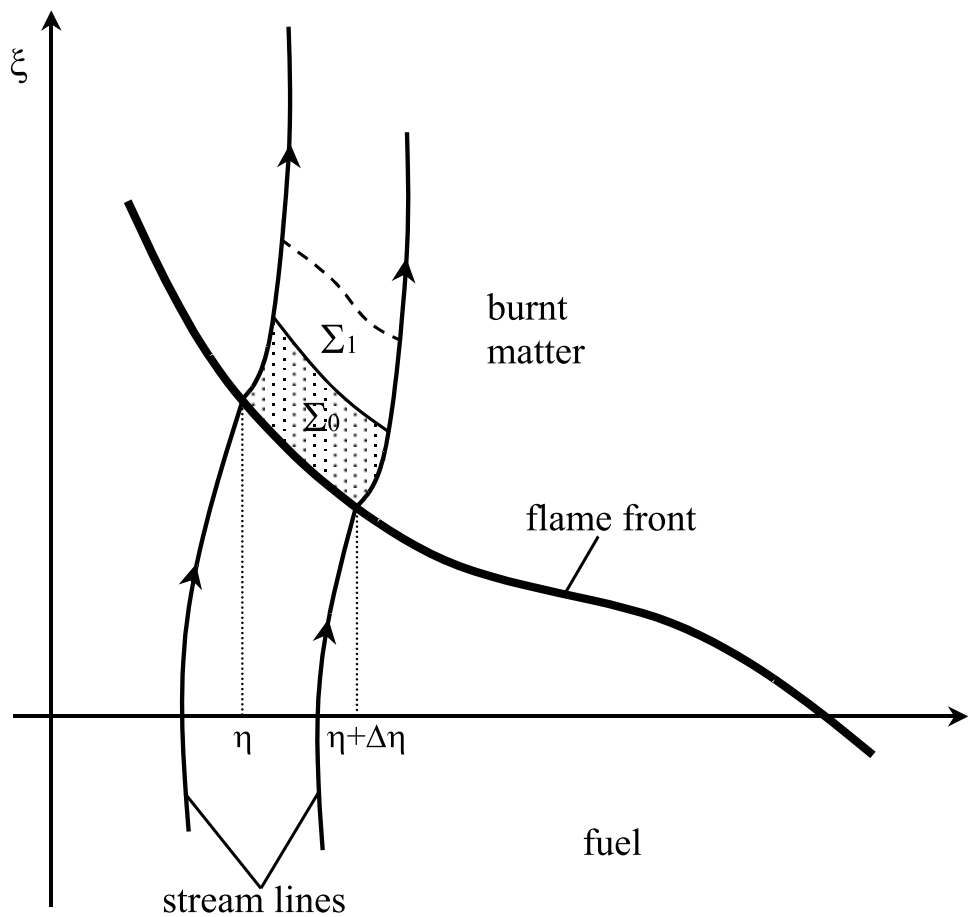}}\caption{Kazakov,
Physics of Fluids} \label{fig1}
\end{figure}

Let us take a couple of stream lines crossing the flame front at
points $(\eta, f(\eta))$ and $(\eta + \Delta\eta, f(\eta +
\Delta\eta))$ (see Fig.~\ref{fig1}). Consider the gas elements
moving between these lines, which cross the front between the time
instants  $t = 0$ and $t = \Delta t.$ During this time interval,
these elements fill a space element $\Sigma_0$ adjacent to the
flame front. For sufficiently small $\Delta\eta, \Delta t,$ the
volume of $\Sigma_0$
$$\Delta s \approx \left|
\begin{array}{cc}
\Delta\eta & f'\Delta\eta \\
w_+ \Delta t & u_+\Delta t \\
\end{array}
\right| = (u_+ - f'w_+)\Delta\eta \Delta t = v^n_+ N \Delta\eta
\Delta t\,,$$ where $$f'\equiv \frac{d f}{d\eta}\,, \quad N\equiv
\sqrt{1 + \left(f'\right)^2}\ ;$$ the subscript``$+$'' means that
the corresponding quantity is evaluated just behind the flame
front, {\it i.e.,} for $\xi = f(\eta) + 0,$ and $v^n_+ = v_{i+}
n_i$ is the normal velocity of the burnt gas, $n_i$ being the unit
vector normal to the flame front (pointing to the burnt matter).
After another time interval of the same duration $\Delta t,$ the
elements move to a space element $\Sigma_1$ adjacent to
$\Sigma_0.$ Since the flow is incompressible, $\Sigma_1$ is of the
same volume as $\Sigma_0.$ Continuing this, the space between the
two stream lines turns out to be divided into an infinite sequence
of $\Sigma$'s of the same volume, adjacent to each other. Thus,
summing over all $\Delta\eta,$ the third term in Eq.~(\ref{vint})
can be written as
\begin{eqnarray}\label{vint1}
- \frac{\varepsilon_{ik}}{2}\partial_k \int\limits_{F} d
l~v^n_+\sigma_{+} K(\eta,\xi,\tilde{\eta})\,,
\end{eqnarray}
\noindent where $F$ denotes the flame front surface (the front
line in our 2D case),
\begin{eqnarray}\label{kernel}
K(\eta,\xi,\tilde{\eta}) &=& \frac{1}{\pi}\lim\limits_{\Delta t\to
0} \sum\limits_{n=0}^{\infty} \ln\left\{(\eta - {\rm
H}(\tilde{\eta},n \Delta t))^2 + (\xi - \Xi(\tilde{\eta},n\Delta
t))^2\right\}^{1/2}\Delta t
\nonumber\\
&=& \frac{1}{\pi}\int\limits_{0}^{+ \infty}d t \ln\left\{(\eta -
{\rm H}(\tilde{\eta},t))^2 + (\xi -
\Xi(\tilde{\eta},t))^2\right\}^{1/2},
\end{eqnarray}
\noindent and $({\rm H}(\tilde{\eta},t),\, \Xi(\tilde{\eta},t))$
trajectory of a particle crossing the point $(\tilde{\eta},
f(\tilde{\eta}))$ at $t = 0.$

Substituting expression (\ref{vint1}) into Eq.~(\ref{vint}) gives
\begin{eqnarray}\label{vint2}
v_i &=& \varepsilon_{ik}\partial_k \int\limits_{\Lambda}d
l_l~\varepsilon_{lm} v_m\frac{\ln r}{2\pi} -
\partial_i\int\limits_{\Lambda}d l_k~v_k\frac{\ln r}{2\pi} -
\frac{\varepsilon_{ik}}{2}\partial_k \int\limits_{F} d
l~v^n_+\sigma_{+} K(\eta,\xi,\tilde{\eta})\,.
\end{eqnarray}
\noindent This representation of the flow velocity downstream will
be used in the next section to determine the structure of the
vortex mode near the flame front.

\section{Structure of the vortex mode}\label{structure}

To determine the flame front dynamics, it is sufficient to know
the flow structure near the flame front. As to the flow upstream,
it is described by Eqs.~(\ref{solup1}), (\ref{solup2}) for all
$\xi < f(\eta).$ Given the solution upstream, velocity components
of the burnt gas at the flame front can be found from the jump
conditions which express the mass and momentum conservation across
the front. On the other hand, these components are required to be
the boundary values (for $\xi = f(\eta) + 0$) of the velocity
field satisfying the flow equations. As was shown in the preceding
section, the latter can be represented in the integral form,
Eq.~(\ref{vint2}). Any velocity field can be arbitrarily
decomposed into a potential ($v_i^p$) and vortex ($v_i^v$) modes:
$$v_i = v_i^p + v_i^v\,.$$ Our strategy below will be to use the integral
representation to determine the near-the-front structure of the
vortex mode described by the last term in Eq.~(\ref{vint2}). As to
$v_i^p,$ the condition of its potentiality, expressed in the form
of a ``dispersion relation'' at the flame front, will eventually
close the system of integro-differential equations at the front.

Equation (\ref{vint2}) reveals the following important fact. Up to
a potential, the value of the vortex mode at a given point
$(\eta,\xi)$ of the flow downstream is determined by only one
point in the range of integration over $(\tilde{\eta},t),$ namely
that satisfying
\begin{eqnarray}\label{p}
{\rm H}(\tilde{\eta},t) = \eta\,,\quad \Xi(\tilde{\eta},t) =
\xi\,.
\end{eqnarray}
\noindent This is, of course, a simple consequence of the Thomson
theorem underlying the above derivation of Eq.~(\ref{vint2}). It
can be verified directly by calculating the curl of the right hand
side of Eq.~(\ref{vint2}): contracting this equation with
$\varepsilon_{im}\partial_m,$ using $\varepsilon_{ik}
\varepsilon_{im} = \delta_{km},$ and taking into account the
relation
\begin{eqnarray}\label{green}
\triangle\ln r = 2\pi\delta^{(2)}(\zeta-\tilde{\zeta})\,,
\end{eqnarray}
\noindent one finds
\begin{eqnarray}\label{twod}
\frac{1}{2}\varepsilon_{im}\partial_m\varepsilon_{ik}\partial_k
\int\limits_{F} d
l~v^n_+\sigma_{+}(\tilde{\eta})K(\eta,\xi,\tilde{\eta}) =
\int\limits_{0}^{+\infty}d t \int\limits_{F} d
l~v^n_+\sigma_{+}(\tilde{\eta}) \delta(\eta - {\rm
H}(\tilde{\eta},t))\delta(\xi - \Xi(\tilde{\eta},t))\,.
\end{eqnarray}
\noindent Since $\delta (x) = 0$ for $x\ne 0,$ the product of
$\delta$-functions picks the point (\ref{p}) out of the whole
range of integration in the right hand side of Eq.~(\ref{twod}).

Now, let us take the observation point $(\xi,\eta)$ sufficiently
close to the flame front, {\it i.e.,} $\xi\approx f(\eta),$
[$\xi>f(\eta)$]. In view of what has just been said, the vortex
component for such points is determined by a contribution coming
from the integration over $\tilde{\eta},t$ near the flame front,
which corresponds to small values of $t.$ Integration over all
other $\tilde{\eta},t$ gives rise to a potential contribution.

The small $t$ contribution to the integral kernel
$K(\eta,\xi,\tilde{\eta})$ can be calculated exactly. For such
$t$'s, one can write
\begin{eqnarray}\label{stream}
{\rm H}(\tilde{\eta},t)\approx \tilde{\eta} +
w_+(\tilde{\eta})t\,,\quad \Xi(\tilde{\eta},t)\approx
f(\tilde{\eta}) + u_+(\tilde{\eta})t\,.
\end{eqnarray}
Let the equality of two fields $\varphi_1(\eta,\xi),\,
\varphi_2(\eta,\xi)$ up to a potential field be denoted as
$\varphi_1\stackrel{\circ}{=} \varphi_2.$ Then, substituting
Eq.~(\ref{stream}) into Eq.~(\ref{kernel}), and integrating gives
\begin{eqnarray}\label{kernel1}
K(\eta,\xi,\tilde{\eta}) &=& \frac{1}{\pi}\int\limits_{0}^{+
\infty}d t \ln\left\{(\eta - {\rm H}(\tilde{\eta},t))^2 + (\xi -
\Xi(\tilde{\eta},t))^2\right\}^{1/2}
\nonumber\\
&\stackrel{\circ}{=}& \frac{1}{\pi}\int\limits_{0}^{t_0}d t
\ln\left\{ v_+^2 t^2 - 2 \bm{r v_+}t + r^2\right\}^{1/2}
\nonumber\\
&=& \frac{1}{\pi}\int \limits_{-\frac{\bm{r v_{+}}}{v_+}}^{v_+ t_0
- \frac{\bm{r v_+}}{v_+}}d t_1 \ln\left\{t_1^2 - \frac{(\bm{r
v_+})^2}{v_+^2} + r^2\right\}^{1/2}
\nonumber\\
&=& \frac{1}{\pi v_+}\left[\sqrt{r^2 - \frac{(\bm{r
v_+})^2}{v_+^2}} \arctan\frac{t_1}{\sqrt{r^2 - \frac{(\bm{r
v_+})^2}{v_+^2}}} \right. \nonumber\\&& \left. - t_1 +
t_1\ln\left\{t_1^2 - \frac{(\bm{r v_+})^2}{v_+^2} +
r^2\right\}^{1/2} \right]_{-\frac{\bm{r v_+}}{v_+}}^{v_+ t_0 -
\frac{\bm{r v_+}}{v_+}}\,.
\end{eqnarray}
\noindent Here  $v_+$ denotes the absolute value of the velocity
field at the flame front, and $t_0$ is assumed small enough to
justify the approximate equations (\ref{stream}).

As we know, the only point in the range of integration over
$\tilde{\eta},t$ that contributes to the vortex mode is the one
satisfying Eq.~(\ref{p}) or, after integrating over $t,$
\begin{eqnarray}\label{point}
[\eta - \tilde{\eta}] u_+(\tilde{\eta}) - [\xi - f(\tilde{\eta})]
w_+(\tilde{\eta}) = 0\,.
\end{eqnarray}
The distance $r$ between this point and the point of observation
tends to zero as the latter approaches to the flame front surface.
Thus, taking $[\xi - f(\eta)]$ small enough, one can make the
ratio $t_0/r$ as large as desired; therefore, the right hand side
of Eq.~(\ref{kernel1}) is $\stackrel{\circ}{=}$
\begin{eqnarray}\label{kernel2}
\frac{r}{\pi v_+} \left\{\sqrt{1 - \left(\frac{\bm{r v_+}}{r
v_+}\right)^2} \left[\frac{\pi}{2} + \arcsin\left(\frac{\bm{r
v_+}}{r v_+}\right) \right] - \frac{\bm{r v_+}}{r v_+} +
\frac{\bm{r v_+}}{r v_+}\ln r \right\} + {\rm TIC}\,,
\end{eqnarray}
\noindent where ``TIC'' stands for ``Terms Independent of the
Coordinates'' $(\eta,\xi).$ Denoting $$\Omega = \frac{\bm{r
v_+}}{r v_+}\,,$$ we finally obtain the following expression for
the integral kernel
\begin{eqnarray}
K(\eta,\xi,\tilde{\eta}) \stackrel{\circ}{=} \frac{r}{\pi v_+}
\left\{\sqrt{1 - \Omega^2} \left(\frac{\pi}{2} +
\arcsin\Omega\right) + \Omega\ln \frac{r}{e}\right\} + {\rm
TIC}\,. \nonumber
\end{eqnarray}

In order to find the vortex mode of the velocity according to
Eq.~(\ref{vint2}) we need to calculate derivatives of $K.$ Using
the relation
\begin{eqnarray}\label{aux1}
\frac{\partial\Omega}{\partial\zeta_i} =
\frac{1}{r}\left(\frac{v_{i+}}{v_+} -
\Omega\frac{r_i}{r}\right)\,, \qquad r_i = \zeta_i -
\tilde{\zeta}_i\,,
\end{eqnarray}
one easily obtains
\begin{eqnarray}\label{kderiv}
\frac{\partial K}{\partial\zeta_i} \stackrel{\circ}{=}
\frac{1}{\pi v_+}\left\{\left(\frac{r_i}{r} -
\Omega\frac{v_{i+}}{v_+}\right) \frac{\pi/2 +
\arcsin\Omega}{\sqrt{1-\Omega^2}} + \frac{v_{i+}}{v_+}\ln
r\right\}\,.
\end{eqnarray}
\noindent Equation (\ref{kderiv}) can be highly simplified.
Consider the quantity
\begin{eqnarray}\label{potential}
\Upsilon_i = \left(\frac{r_i}{r} - \Omega\frac{v_{i+}}{v_+}\right)
\frac{\arcsin\Omega - \pi/2}{\sqrt{1-\Omega^2}} +
\frac{v_{i+}}{v_+}\ln r\,.
\end{eqnarray}
\noindent Let us evaluate $\partial_i\Upsilon_i\,.$ First, we
calculate
\begin{eqnarray}\label{aux2}
\frac{\partial}{\partial\zeta_i} \left(\frac{r_i}{r} -
\Omega\frac{v_{i+}}{v_+}\right) = \left(\frac{\partial_i r_i}{r} -
r_i\frac{r_i}{r^3}\right) - \frac{\partial\Omega}{\partial
r_i}\frac{v_{i+}}{v_+} = \frac{\Omega^2}{r}\,,
\end{eqnarray}
\begin{eqnarray}\label{aux3}
\frac{\partial}{\partial\zeta_i} \left(\frac{v_{i+}}{v_+}\ln
r\right) = \frac{v_{i+}}{v_+}\frac{r_i}{r^2} = \frac{\Omega}{r}\,.
\end{eqnarray}
\noindent Second, we note that the vector
$$\beta_i = \left(\frac{r_i}{r} - \Omega\frac{v_{i+}}{v_+}\right)
\frac{1}{\sqrt{1-\Omega^2}}$$ satisfies
$$\beta_i\beta_i = 1, \qquad \beta_i v_{i+} = 0\,,$$
{\it i.e.,} $\beta_i$ is the unit vector orthogonal to ${\bf
v}_+\,.$ In addition to that, $\beta_i$ changes its sign at the
point defined by Eq.~(\ref{point}). Therefore, the derivative of
$\beta_i,$ entering $\Upsilon,$ contains a term with the Dirac
$\delta$-function. However, this term is multiplied by
$(\arcsin\Omega - \pi/2)$ which turns into zero together with the
argument of the $\delta$-function. Therefore, the product of the
additional term with $(\arcsin\Omega - \pi/2)$ is to be set zero,
in the sense of distributions. Thus, using Eqs.~(\ref{aux1}),
(\ref{aux2}), (\ref{aux3}) one finds
\begin{eqnarray}
\frac{\partial \Upsilon_i}{\partial\zeta_i} =
\frac{\Omega^2}{r}\frac{\arcsin\Omega - \pi/2} {\sqrt{1 -
\Omega^2}} + \left(\frac{r_i}{r} -
\Omega\frac{v_{i+}}{v_+}\right)\left[\frac{1}{1 - \Omega^2} +
\Omega\frac{\arcsin\Omega - \pi/2}{(1 - \Omega^2)^{3/2}}\right]
\frac{\partial\Omega}{\partial\zeta_i} + \frac{\Omega}{r} \equiv
0\,. \nonumber
\end{eqnarray}
\noindent We conclude that $\Upsilon_i$ gives rise to a pure
potential. A similar calculation shows that also
\begin{eqnarray}\label{divfree}
\varepsilon_{ik}\frac{\partial \Upsilon_i}{\partial\zeta_k} \equiv
0\,.
\end{eqnarray} \noindent
Therefore, we can rewrite Eq.~(\ref{kderiv}) as
\begin{eqnarray}\label{kderiv1}
\frac{\partial K}{\partial\zeta_i} \stackrel{\circ}{=}
\frac{1}{v_+}\left(\frac{r_i}{r} - \Omega\frac{v_{i+}}{v_+}\right)
\frac{1}{\sqrt{1-\Omega^2}} = \frac{\beta_i}{v_+}\,.
\end{eqnarray}
\noindent Finally, substituting this result into
Eq.~(\ref{vint2}), noting that the vector
$\varepsilon_{ki}\beta_k$ is the unit vector parallel to $v_{i+}$
if $\varepsilon_{ik}r_i v_{k+}>0,$ and antiparallel in the
opposite case, we obtain the following expression for the vortex
component of the gas velocity downstream near the flame front
\begin{eqnarray}\label{vintf}
v^{v}_i = \int\limits_{F} d l~\chi(\varepsilon_{pq}r_p v_{q+})
\frac{v^n_+\sigma_{+}v_{i+}}{2 v^2_+}\,.
\end{eqnarray}
\noindent Having written the exact equality in Eq.~(\ref{vintf})
we take this equation as the {\it definition} of the vortex mode.
As a useful check, it is verified in appendix~\ref{appa} that the
obtained expression for $v^v_i$ satisfies
$$\left(\partial u^v/\partial\eta -
\partial w^v/\partial\xi\right)_+ \equiv \sigma_+\,.$$

It remains only to make the following comment in connection with
the obtained expression for the vortex mode. As is clear from its
derivation, Eq.~(\ref{vintf}) is applicable to unbounded as well
as bounded flames. In the former case, however, the improper
integral on the right of this equation is undefined, because
integration over the infinite ``tails'' of the flame front around
the point satisfying Eq.~(\ref{point}) gives rise to a potential
contribution which is formally divergent. In the case of periodic
flames which are of our main concern (see the beginning of
Sec.~\ref{flowequations}) this complication can be easily overcome
if we specify that the integral is to be understood as
\begin{eqnarray}\label{rule}
\int\limits_{F} d \eta \cdots = \lim\limits_{\mathbb{N}\ni n
\to\infty} \int\limits_{-2bn}^{+2bn} d \eta \cdots\,,
\end{eqnarray}
\noindent so that the contributions of the tails cancel each other
exactly. We prefer to work with an infinite $\eta$-interval,
rather than $\eta\in [-b,b],$ because of the reasons that will be
clear in Sec.~\ref{dispersion}.

\section{Closed description of stationary flames}\label{closed}

After we have determined the near-the-front structure of the
vortex component of the gas velocity downstream, we can write down
a closed system of equations governing the stationary flame
propagation. As was explained in the Introduction, the term
``closed'' means that these equations relate only quantities
defined on the flame front surface, without any reference to the
flow dynamics in the bulk. This system consists of the jump
conditions for the velocity components at the front, and the
so-called evolution equation that gives the local normal fuel
velocity at the front as a function of the front curvature. These
equations (except for the evolution equation) are consequences of
the mass and momentum conservation across the flame front. In
Sec.~\ref{gen}, we obtain a closed system in the most general
form, without specifying the form of the jump conditions, and then
apply it to the case of zero thickness flames in Sec.~\ref{zero}.

\subsection{General formulation}\label{gen}

First of all, we need to find the ``on-shell'' expression for the
vortex component,  i.e. its limiting form for $\xi\to f(\eta) +
0.$ For this purpose we note that in this limit,
$\chi(\varepsilon_{ik}r_i v_{k+})\to \chi(\eta - \tilde{\eta}),$
therefore, Eq.~(\ref{vintf}) gives
\begin{eqnarray}\label{vinton}
v^{v}_{i+} = \int\limits_{F} d l~\chi(\eta - \tilde{\eta})
\frac{v^n_+\sigma_{+}v_{i+}}{2 v^2_+}\,.
\end{eqnarray}
\noindent Let us denote the jump $[v_{i+}(\eta) - v_{i-}(\eta)]$
of the gas velocity $v_i$ across the flame front as $[v_i].$ Here
$v_{i\pm}(\eta) \equiv v_{i}(\eta,f(\eta)\pm 0).$ Then we can
write

\begin{eqnarray}\label{general}
v_{i-} + [v_i] = v^p_{i+} + \int\limits_{F} d l~\chi(\eta -
\tilde{\eta})\sigma_{+} \frac{(v^n_- + [v^n]) (v_{i-} + [v_i])}{2
(v_-^2 + [v^2])}\,.
\end{eqnarray}
\noindent In every finite order of the asymptotic expansion with
respect to the flame front thickness, the jumps $[v_i]$ (as well
as $\sigma_+$) are quasi-local functionals of the fuel velocity at
the flame front, and of the flame front shape (i.e., depend on
their values and the values of their derivatives of finite order
in a given point). Two equations (\ref{general}), together with
Eq.~(\ref{solup2}) and the evolution equation, $v^n_- = v^n_-(f),$
form a system of four equations for the five functions
$v_{i-}(\eta), v^p_{i+}(\eta),$ and $f(\eta).$ To close this
system, we need an equation expressing potentiality of the field
$v^p_{i},$ to be derived in the next section.

\subsubsection{A dispersion relation for the potential
mode}\label{dispersion}

By the construction, divergence of the last term in
Eq.~(\ref{vint2}), describing the vortex mode, is zero identically
in the entire downstream region. It is not difficult to see that
it retains this property after the number of simplifications we
have made in the course of derivation of the final expression
(\ref{vintf}). Note, first of all, that any operating with the
integral kernel $K(\eta,\xi,\tilde{\eta})$ itself {\it before}
differentiation with respect to $\zeta_i$ cannot break this
property. On the other hand, Eq.~(\ref{divfree}) shows that
divergence of the only term, namely $\Upsilon_i,$ that have been
omitted {\it after} this differentiation is zero identically. In
view of the equation ${\rm div}\,\bm{v} = 0,$ one concludes that
the potential mode of the velocity field, $\bm{v}^{p},$ satisfies
${\rm div}\,\bm{v}^{p} = 0$ downstream, too. Equations ${\rm
rot}\,\bm{v}^{p} = 0,$ ${\rm div}\,\bm{v}^{p} = 0$ allow us to
introduce the potential, $\phi,$ and the stream function, $\psi,$
according to
\begin{eqnarray}\label{phipsi}
u^p = \frac{\partial\phi}{\partial\xi} =
\frac{\partial\psi}{\partial\eta}\,, \qquad w^p =
\frac{\partial\phi}{\partial\eta} = -
\frac{\partial\psi}{\partial\xi}\,.
\end{eqnarray}
\noindent These relations imply that the combination $\Phi = \psi
+ i\phi$ is an analytical function of the complex variable $z =
\eta + i\xi,$ and therefore so is its derivative $d\Phi/d z \equiv
\omega^p = u^p + iw^p.$ Then, using the Cauchy theorem, we can
write
\begin{eqnarray}\label{cauchy}
\omega^p(z) = \frac{1}{2\pi i}\oint\limits_{\Lambda} d\tilde{z}
\frac{\omega^p(\tilde{z})}{\tilde{z} - z}\,,
\end{eqnarray}
\noindent where $z\in \Sigma,$ and it is assumed that $\tilde{z}$
runs $\Lambda$ counterclockwise. In the course of derivation of
the expression (\ref{vintf}) for the vortex mode, we have been
systematically omitting potential contributions. By the
construction, these contributions are proportional to the integral
kernel of the Laplace operator [Cf. Eq.~(\ref{vint2})], and
therefore, they generally diverge logarithmically at infinity [as,
for instance, the last term in the expression (\ref{potential})].
Hence, the integral over the part $\Lambda \diagdown F$ of the
boundary of $\Sigma$ in Eq.~(\ref{cauchy}) is formally an infinite
constant, while the improper integral over $F$ is undefined. To
avoid appearance of such divergent integrals, we will work below
with the velocity derivative $d\omega^p/dz,$ instead of
$\omega^p.$ Then Eq.~(\ref{cauchy}) is replaced by
\begin{eqnarray}\label{cauchy1}
\frac{d\omega^p}{d z}(z) = \frac{1}{2\pi i}\oint\limits_{\Lambda}
\frac{d\tilde{z}}{\tilde{z} - z}
\frac{d\omega^p}{dz}(\tilde{z})\,.
\end{eqnarray}
\noindent Let us show that this identity can be rewritten as a
dispersion relation for $d\omega/dz$ taken at the flame front. For
this purpose, let us choose the contour of integration consisting
of a large semicircle of radius $R,$ its center being at the point
$z_0 = 0 + i f(0),$ and of the part of the front $F,$ indented by
the circle. Then in the limit $z\to F,$ Eq.~(\ref{cauchy1}) takes
the form
\begin{eqnarray}\label{cauchy3}
\frac{d\omega^p}{d z}(z_+) = \frac{1}{\pi i}\dashint\limits_{F}
\frac{d\tilde{z}}{\tilde{z} - z_+} \frac{d\omega^p}{dz}(\tilde{z})
+ \frac{1}{\pi}\int\limits_{0}^{\pi} d\varphi
\lim\limits_{R\to\infty}\frac{d\omega^p}{dz}\left(R
e^{i\varphi}\right)\,.
\end{eqnarray}
\noindent Our choice of the contour implies that the first
integral on the right hand side of this equation is understood as
\begin{eqnarray}\label{rulec}
\int\limits_{F}d\tilde{z}\cdots = \lim\limits_{\mathbb{N}\ni n \to
\infty}\int\limits_{z_0-2bn}^{z_0+2bn}d\tilde{z}\cdots\,.
\end{eqnarray}
\noindent According to what has been said about behavior of the
potential mode at infinity, $$\frac{d\omega^p}{dz}\left(R
e^{i\varphi}\right) \to 0 \quad {\rm as}\quad R\to \infty \quad
{\rm for\ \   all} \quad \varphi \in (0,\pi).$$

Next, differentiating Eq.~(\ref{vinton}) with respect to $\eta,$
taking into account the relation
\begin{eqnarray}\label{deltad}
\frac{d\chi(x)}{d x} = 2\delta(x)\,,
\end{eqnarray}
\noindent and  performing the $\tilde{\eta}$-integration yields
\begin{eqnarray}\label{vintond}
\left(v^{v}_{i+}\right)' =  \frac{N v^n_+\sigma_{+}v_{i+}}{
v^2_+}\,.
\end{eqnarray}
\noindent On the other hand, one can write, in view of analyticity
of $\omega^p,$
$$\left(\omega^p_+\right)' =
\left(\frac{\partial\omega^p}{\partial\eta}\right)_+ +
f'\left(\frac{\partial\omega^p}{\partial\xi}\right)_+ = \left(1 +
i f'\right)\left(\frac{d\omega^p}{d z}\right)_+\,,$$ or,
$$\left(\frac{d\omega^p}{d z}\right)_+ = \frac{1}{\left(1 +
i f'\right)}\left[\left(\omega_+\right)' - \frac{N
v^n_+\sigma_{+}\omega_+}{ v^2_+}\right]\,, \qquad \omega = u + i
w\,.$$ The right hand side of this equation is explicitly
periodic. Hence, the first integral in Eq.~(\ref{cauchy3}) is
well-defined by the rule (\ref{rulec}), so this equation becomes
\begin{eqnarray}\label{cauchy4}
\left(\frac{d\omega^p}{d z}\right)_+ = \frac{1}{\pi
i}\dashint\limits_{F} \frac{d\tilde{z}}{\tilde{z} -
z_+}\left(\frac{d\omega^p}{d z}\right)_+ \,.
\end{eqnarray}
\noindent A relation similar to Eq.~(\ref{cauchy4}) can be written
for the gas velocity upstream:
\begin{eqnarray}\label{cauchy5}
\left(\frac{d\omega}{d z}\right)_- = - \frac{1}{\pi
i}\dashint\limits_{F} \frac{d\tilde{z}}{\tilde{z} -
z_-}\left(\frac{d\omega}{d z}\right)_- \,.
\end{eqnarray}

We have avoided appearance of potentially divergent quantities in
the dispersion relation for the gas flow downstream at the cost of
increasing its differential order by one$.^{13}$ Otherwise, we
would have had to carry out all intermediate calculations for a
finite domain $\Sigma,$ and to require all divergences to cancel
in the final equation for the flame front position in the limit of
infinite front length. In both cases, the finite constant in this
equation (an integration constant in the first case, or a finite
remainder after the cancellation in the second) is fixed by the
normalization condition
\begin{eqnarray}\label{lvcondition}
V = \frac{1}{2b}\int\limits_{-b}^{+b} d\eta~N\,.
\end{eqnarray}
\noindent

\subsubsection{The integro-differential relation between $\omega_-$
and $f.$}\label{mainrel}

We are now in a position to write down the main
integro-differential equation relating the values of the gas
velocities at the flame front with the flame front position. To
this end, we first differentiate Eq.~(\ref{general}) with respect
to $\eta$ doing the integral as before, and rewrite the result in
the complex form:
\begin{eqnarray}\label{generalc}
\left(\omega_-\right)' + [\omega]' = \left(\omega^p_+\right)' +
N\sigma_{+} \frac{(v^n_- + [v^n]) (\omega_- + [\omega])}{(v_-^2 +
[v^2])}\,.
\end{eqnarray}
\noindent Acting on this equation by the operator $$\left(1 + i
f'\right)\left(1 - \frac{1}{\pi i}\dashint\limits_{F} \frac{d
\tilde{z}}{\tilde{z} - z_+}\right)\left(1 + i f'\right)^{-1}\,,$$
taking into account Eqs.~(\ref{cauchy4}), (\ref{cauchy5}), and the
relations
$$\left(\omega^p_+\right)' = \left(1 + i f'\right)
\left(\frac{d\omega^p}{d z}\right)_+\,, \qquad
\left(\omega_-\right)' = \left(1 + i f'\right)
\left(\frac{d\omega}{d z}\right)_- $$ gives
\begin{eqnarray}\label{generalc1}&&
2\left(\omega_-\right)' + \left(1 +
i\hat{\EuScript{H}}\right)\left\{[\omega]' - N\sigma_{+}
\frac{(v^n_- + [v^n]) (\omega_- + [\omega])}{(v_-^2 +
[v^2])}\right\} = 0\,,
\end{eqnarray}
\noindent where $$\hat{\EuScript{H}} = \frac{1 + i
f'}{\pi}\dashint\limits_{F} \frac{d \tilde{\eta}}{\tilde{z} -
z_+}\,.$$ Written longhand, the action of the operator
$\hat{\EuScript{H}}$ on a function $a(\eta)$ is
\begin{eqnarray}
\left(\hat{\EuScript{H}}a\right)(\eta) = \frac{1 + i
f'(\eta)}{\pi}~\dashint\limits_{-\infty}^{+\infty}
d\tilde{\eta}~\frac{a(\tilde{\eta})}{\tilde{\eta} - \eta +
i[f(\tilde{\eta}) - f(\eta)]}\,.
\end{eqnarray}
\noindent $\hat{\EuScript{H}}$ has properties similar to the
Hilbert operator $\hat{H}.$ In particular, it is shown in
appendix~\ref{b1} that
\begin{eqnarray}\label{pr}
\hat{\EuScript{H}}^2 = - 1\,.
\end{eqnarray}
\noindent In terms of $\hat{\EuScript{H}},$ $\omega_-,$ the
dispersion relation (\ref{cauchy5}) takes a more compact form
\begin{eqnarray}\label{ch}
\left(1 - i \hat{\EuScript{H}}\right)\left(\omega_-\right)' = 0\,.
\end{eqnarray}
\noindent In view of the property (\ref{pr}), this relation is
fulfilled by any $\omega_-$ satisfying Eq.~(\ref{generalc1}).

Equation (\ref{generalc1}) is the main integro-differential
relation between the fuel velocity at the flame front, and the
flame front position. It can also be rewritten in terms of
$\omega_+$ as
\begin{eqnarray}\label{generalc2}
2\left(\omega_+\right)' - \left\{[\omega]' + N\sigma_{+}
\frac{v^n_+ \omega_+}{v_+^2}\right\} + i\hat{\EuScript{H}}
\left\{[\omega]' - N\sigma_{+} \frac{v^n_+\omega_+}{v_+^2}\right\}
= 0\,.
\end{eqnarray}
\noindent Together with the jump conditions and the evolution
equation which has the general form
\begin{eqnarray}\label{evolutiongen}
v^n_- = 1 + S(u_-,w_-,f')\,,
\end{eqnarray}
\noindent where $S$ is a quasi-local functional of its arguments,
proportional to the flame front thickness, Eq.~(\ref{generalc1})
provides a closed description of stationary flames.

\subsection{Equation (\ref{generalc1}) in lowest orders of
the $\varepsilon$-expansion}

In this section, the general results obtained in the preceding
section will be applied to thin flames in the first two orders of
the asymptotic expansion with respect to the flame front thickness
$\varepsilon = L_{\rm f}/R.$

\subsubsection{Zero-thickness flames}\label{zero}

For zero-thickness flames, the jump conditions for the velocity
components have the form
\begin{eqnarray}\label{jump1}
[u] &=& \frac{\theta - 1}{N}\ ,\\
\label{jump2} [w] &=& - f'\frac{\theta - 1}{N}\,,
\end{eqnarray}
\noindent and hence,
\begin{eqnarray}\label{jump1c}
[\omega] &=& (\theta - 1)\frac{1 - i f'}{N}\ ,
\end{eqnarray}
\noindent
while the evolution equation
\begin{eqnarray}\label{evolution}
v^n_- = 1\,.
\end{eqnarray}
\noindent We see that the jumps are velocity-independent, and
$S\equiv 0.$ Also, it follows from these equations that
$$ [v^n] = \theta - 1\,, \quad [v^2] = \theta^2 - 1\,.$$

It remains only to calculate the value of the vorticity at the
flame front, as a function of the fuel velocity. This can be
done$^{2}$ directly using the flow equations
(\ref{flow1}),(\ref{flow2}). With the help of Eqs.~(5.32) and
(6.15) of Ref.~2, the jump of the vorticity across the flame front
can be written, in the 2D stationary case, as
\begin{eqnarray}\label{vort1}&&
[\sigma] = - \frac{\theta - 1}{\theta N} \left(\hat{D}w_{-} +
f'\hat{D}u_{-} + \frac{1}{N}\hat{D}f'\right),
\end{eqnarray}
\noindent where
\begin{eqnarray}\label{operator}
\hat{D}\equiv \left(w_{-} + \frac{f'}{N}\right)\frac{d}{d\eta}\,.
\end{eqnarray}
Differentiating the evolution equation written in the form
\begin{eqnarray}\label{evolution1} u_- - f' w_- = N\,,
\end{eqnarray}
\noindent
and writing Eq.~(\ref{vort1}) longhand, expression in
the  brackets can be considerably simplified
\begin{eqnarray}\label{vort2}&&
\hat{D}w_{-} + f'\hat{D}u_{-} + \frac{1}{N}\hat{D}f'\equiv
w_{-}'w_{-} + \frac{(f'w_{-})'}{N} + \frac{(f')^2 u_{-}'}{N} +
f'u_{-}'w_{-} + \frac{N'}{N} \nonumber\\&&
 = \frac{(w_{-}^2)'}{2} + \frac{(u_{-} - N)'}{N}
+ \frac{(N^2 - 1) u_{-}'}{N} + u_{-}'(u_{-} - N) + \frac{N'}{N} =
u_{-}'u_{-} + w_{-}'w_{-}\,.
\end{eqnarray}
\noindent Since the flow is potential upstream, we obtain the
following expression for the vorticity just behind the flame
front$^{14}$
\begin{eqnarray}\label{vort3}&&
\sigma_{+} = - \frac{\theta - 1}{\theta N}(u_{-}'u_{-} +
w_{-}'w_{-})\,.
\end{eqnarray}
\noindent Substituting these expressions into
Eq.~(\ref{generalc1}) yields
\begin{eqnarray}\label{main}
\left(\omega_-\right)' + \frac{\theta - 1}{2}\left(1
+i\hat{\EuScript{H}}\right)\left\{\left(\frac{1 - i f'}{N}\right)'
+ \frac{1}{2}\left(\omega_- + (\theta - 1)\frac{1 - i
f'}{N}\right)\left[\ln(|\omega_-|^2 + \theta^2 -
1)\right]'\right\} &&\hskip-0,4cm = 0\,.\nonumber\\&&
\end{eqnarray}
\noindent Equations (\ref{evolution1}) and (\ref{main}) provide
the closed description of stationary zero-thickness flames in the
most convenient form. Account of the heat conduction -- species
diffusion processes inside the thin flame front changes the right
hand side of this equation to $O(\varepsilon).$ This modification
will be considered in the next section.

\subsubsection{Account of the transport processes in the
linear approximation}\label{ltransport}

An equation relating $\omega_-$ with $f$ can be obtained also in
the case of flames of nonzero thickness following the lines of the
above derivation of Eq.~(\ref{main}). For instance, in the first
order in the flame front thickness $\varepsilon,$ one has to use
in the general Eq.~(\ref{generalc1}) the jump conditions and the
evolution equation derived in Ref.~2. The resulting equation turns
out to be much more complicated than Eq.~(\ref{main}). However,
the main purpose of taking into account the inner structure of the
flame front is to provide a short wavelength cutoff for unstable
flame perturbations, which is necessary for the very existence of
stationary configurations of curved flames. On the other hand, for
many purposes it is sufficient to consider the transport processes
in the linear approximation, while full account of the nonlinear
coupling of these processes to flame hydrodynamics is of minor
importance in practice. Therefore, we can generalize
Eq.~(\ref{main}) to the case of flames of nonzero thickness almost
without calculation using the results of the linear theory of
flame front instability. In the linear approximation, $\omega_- =
1,$ hence, the left hand side of Eq.~(\ref{main}) reduces to
$$\frac{\theta - 1}{2}\left(\hat{H} - i\right)f''\,.$$
On the other hand, omitting the time derivatives in the linear
equation$^{3}$ describing evolution of the front perturbations
gives formally
$$\hat{H}f' = \frac{\lambda_c}{2\pi}f''\,,$$ where
$\lambda_c = L_c/R=O(\varepsilon)$ is the dimensionless cutoff
wavelength. We conclude that the desired generalization of
Eq.~(\ref{main}) reads$^{15}$
\begin{eqnarray}\label{main1}&&
\left(\omega_-\right)' + \frac{\theta - 1}{2}\left(1
+i\hat{\EuScript{H}}\right)\left\{\left(\frac{1 - i f'}{N}\right)'
+ \frac{1}{2}\left(\omega_- + (\theta - 1)\frac{1 - i
f'}{N}\right)\left[\ln(|\omega_-|^2 + \theta^2 -
1)\right]'\right\} \nonumber\\&& = \frac{\theta
-1}{2}\frac{\lambda_c}{2\pi}\left(1 +
i\hat{\EuScript{H}}\right)f'''\,.
\end{eqnarray}
\noindent Equivalently, the main system of equations can be
rewritten in terms of $w_+, u_+:$
\begin{eqnarray}\label{main1alt}&&
\hspace{-1cm}\left(\omega_+\right)' + \frac{\theta - 1}{2}\left(-
1 +i\hat{\EuScript{H}}\right)\left(\frac{1 - i f'}{N}\right)' +
\frac{\theta - 1}{2}\left(1
+i\hat{\EuScript{H}}\right)\left(\frac{\omega_+|\omega_+|'}
{|\omega_+|} - \frac{\lambda_c}{2\pi}f'''\right) = 0\,,
\\&&\hspace{-1cm} u_+ - f'w_+ = \theta N\,.\nonumber
\end{eqnarray}
\noindent We have written $\hat{\EuScript{H}}$ instead of
$\hat{H}$ in the $\varepsilon$-terms in order to formally preserve
the dispersion relation (\ref{ch}). However, it should be kept in
mind that by the construction, the $\varepsilon$-term is to be
treated linearized.

Most probably, Eq.~(\ref{main1}) can be solved in full generality
only numerically. However, it admits analytical solutions in
lowest orders of the asymptotic expansion for $\theta \to 1,$
considered below.

\section{The small $(\theta - 1)$ expansion}\label{smallt}

Although our main concern in this paper is the non-perturbative
description of flames with arbitrary $\theta,$ it is of some
interest to apply the above results to the case of weak gas
expansion. One reason is that this case allows considerable
simplification of Eqs.~(\ref{evolution1}), (\ref{main1}) which can
be reduced to a single equation for the flame position. Another is
that this is a good place to illustrate the role of the relation
(\ref{lvcondition}) in our approach. The well-known subtle point
of the analytical theory of nonlinear flame propagation is to
ensure that the constant $V$ and the function $f(\eta),$ which
play the role of the eigenvalue and the eigenfunction of the
nonlinear equation for the front position, respectively, satisfy
this obvious condition.

In the present formulation, this problem does not arise at all. To
see this, it is sufficient to note that the constant $V$ does not
appear at all either in Eq.~(\ref{main1}), or in the evolution
equation (\ref{evolution1}). Therefore, Eq.~(\ref{lvcondition}) is
to be considered simply as a relation that allows one to express
the flame velocity through the constant of integration $C$ [Cf.
discussion in the end of Sec.~\ref{dispersion}]. In lowest orders
of the small $(\theta - 1)$ expansion, this connection between $V$
and $C$ can be followed out in detail. It will be shown in
Sec.~\ref{3order} that in the first post-Sivashinsky
approximation, Eqs.~(\ref{evolution1}), (\ref{main1}) reduce to
the Cambray-Joulin$^{16}$ version of the well-known
Sivashinsky-Clavin equation$.^{4}$ In Sec.~\ref{4order}, an
equation for the flame front position will be obtained in the
second post-Sivashinsky approximation, which represents a
corrected version of the equation obtained by Kazakov and
Liberman$^{5-6}$ (called there the fourth order equation).

\subsection{The first post-Sivashinsky approximation}\label{3order}

\subsubsection{Equation for the flame front position}

The form of the $\varepsilon$-term in Eq.~(\ref{main1}) shows that
the cutoff wavelength $\lambda_c \sim 1/\alpha,$ $\alpha \equiv
\theta - 1.$ This implies that $f' = O(\alpha),$ and that
differentiation of a flow variable with respect to $\eta$
increases its order by one. Furthermore, it follows from
Eq.~(\ref{evolution1}) that $u_- = 1 + O(\alpha^2),$ since $V =
O(\alpha^2).$ Then Eq.~(\ref{solup2}) tells us that also $w_- =
O(\alpha^2).$ To carry out expansion of Eq.~(\ref{main1}), we need
also to find the asymptotic action of the operator
$\hat{\EuScript{H}}.$ A general procedure of consistent asymptotic
treatment of this operator is formulated in appendix~\ref{b2}. At
the third order in $\alpha,$ according to Eq.~(\ref{apb12}),
$\hat{\EuScript{H}}$ acts on the expression in the curly brackets
on the left hand side of Eq.~(\ref{main1}) as the Hilbert
operator. Thus, within accuracy of the fourth order,
Eq.~(\ref{main1}) can be rewritten as
\begin{eqnarray}
\left(\omega_-\right)' + \frac{\theta - 1}{2}\left(1 +
i\hat{H}\right)\left(\frac{1 - if'}{N} +
\frac{|\omega_-|^2}{2\theta} - \frac{\lambda_c}{2\pi} f''\right)'
= 0\,.\nonumber
\end{eqnarray}
\noindent Expanding the last term in this equation, noting that
with the required accuracy $|\omega_-|^2$ in the parentheses can
be taken to be
\begin{eqnarray}\label{omega3}
|\omega_-|^2 = 1 + \left(f'\right)^2\,,
\end{eqnarray}
\noindent and integrating gives
\begin{eqnarray}\label{main13}
\omega_- - \frac{\theta - 1}{2}\left(1 + i\hat{H}\right)\left(i f'
+ \frac{\lambda_c}{2\pi} f''\right) = C\,,
\end{eqnarray}
\noindent where $C = C_1 + i C_2$ is the constant of integration.
Extracting the real and imaginary parts of this equation, we
obtain
\begin{eqnarray}
u_- &-& \frac{\theta - 1}{2}\left(- \hat{H} f' +
\frac{\lambda_c}{2\pi} f''\right) = C_1\,, \label{u}\\
w_- &-& \frac{\theta - 1}{2}\left(f' + \frac{\lambda_c}{2\pi}
\hat{H}f''\right) = C_2\,.\label{w}\label{wthird}
\end{eqnarray}
\noindent Integrating Eq.~(\ref{w}) over $[-b,+b],$ and taking
into account periodicity of the function $f(\eta)$ shows that $C_2
= 0,$ because
$$\int\limits_{-b}^{+b}d\eta~w_-(\eta) = 0$$ in view of
the definition (\ref{antisym}). Next, multiplying Eq.~(\ref{w}) by
$f',$ subtracting it from Eq.~(\ref{u}), and using the evolution
equation (\ref{evolution1}), we arrive at the single equation for
the flame front position$^{17}$
\begin{eqnarray}\label{3ordereq}
\frac{\theta}{2}\left(f'\right)^2 = \frac{\theta - 1}{2}\left(-
\hat{H}f' + \frac{\lambda_c}{2\pi}f''\right) + C_1 - 1\,.
\end{eqnarray}
\noindent The constant $C_1$ can be expressed through the flame
velocity $V$ using the normalization condition
(\ref{lvcondition}). Namely, integrating Eq.~(\ref{3ordereq}) over
$[-b,+b],$ and applying this condition gives $$\theta \left(V -
1\right) = C_1 - 1\,,$$ thus bringing the third order equation to
the form
\begin{eqnarray}\label{3ordereqf}
- \theta W + \frac{\theta}{2}\left(f'\right)^2 = \frac{\theta -
1}{2}\left(- \hat{H}f' + \frac{\lambda_c}{2\pi}f''\right) \,,
\end{eqnarray}
\noindent where $W \equiv V - 1$ is the flame velocity increase
due to the front curvature. This equation exactly coincides with
the stationary version of the corrected Sivashinsky-Clavin
equation (obtained by setting $\phi_t = - W,$ and restoring the
$\varepsilon$-term in Eq.~(10) of Ref.~16).

\subsubsection{Solution of the equation (\ref{3ordereqf})}
\label{ansol}

The third order equation (\ref{3ordereqf}) is of the same
functional structure as the Sivashinsky equation$.^{1}$ Therefore,
it can be solved analytically using the method of pole
decomposition$.^{18,19}$ We look for a $2 b$-periodic solution of
Eq.~(\ref{3ordereqf}) in the form
\begin{eqnarray}\label{anzats} f(\eta) = A \sum_{k = 1}^{2 P}
\ln\sin\left[\frac{\pi}{2 b}(\eta - \eta_k)\right].
\end{eqnarray}
\noindent The amplitude $A$ and the complex poles $\eta_k,$ $k =
1,...,2P$ are to be determined substituting this anzats into
Eq.~(\ref{3ordereqf}). Since the function $f(\eta)$ is real for
real $\eta,$ the poles come in conjugate pairs; $P$ is the number
of the pole pairs.

Using the formulas (see Refs.~18,19 for more detail)
\begin{eqnarray}&&
\hat{H}f' = - \frac{\pi A}{2 b}\sum_{k = 1}^{2 P}\left\{1 + i~{\rm
 sign}({\rm Im}~\eta_k)\cot\left[\frac{\pi}{2 b}(\eta -
\eta_k)\right]\right\}, ~{\rm sign}(x) \equiv
\frac{x}{|x|}\,,\nonumber\\&& \cot x \cot y = -1 + \cot(x - y
)(\cot y - \cot x)\,, \nonumber
\end{eqnarray}
\noindent it is easily verified that Eq.~(\ref{3ordereqf}) is
satisfied by $f(\eta)$ taken in the form of Eq.~(\ref{anzats}),
provided that
\begin{eqnarray}\label{solution1}
A &=& - \frac{\lambda_c}{2\pi}\frac{\theta - 1 }{\theta}\,,
\nonumber\\ W &=& \frac{(\theta - 1)^2}{2\theta^2
}\frac{P\lambda_c}{2 b}\left(1 - \frac{P\lambda_c}{2 b}\right),
\end{eqnarray}
\noindent and the poles $\eta_k,$ $k = 1,...,2P\,,$ satisfy the
following set of equations
\begin{eqnarray}&&\label{solution2}
i~{\rm sign}({\rm Im}~\eta_k) + \frac{\lambda_c}{2
b}\sum\limits_{\genfrac{}{}{0pt}{}{m = 1}{m\ne k}}^{2
P}\cot\left[\frac{\pi}{2 b}(\eta_k - \eta_m)\right] = 0, ~k =
1,...,2P\,.
\end{eqnarray}
\noindent It is seen from Eq.~(\ref{solution1}) that the found
solution (\ref{anzats}) is not unique: there is a number of
solutions corresponding to different numbers $P$ of poles. To
identify the physical ones, a stability analysis of the solutions
is required which, of course, cannot be carried out in the
framework of the stationary theory. However, as we have mentioned
above, the functional structure of Eq.~(\ref{3ordereqf}) is very
similar to that of the stationary Sivashinsky equation. Under
assumption that the non-stationary versions of these equations are
also similar, the stability analysis performed in Ref.~20 will be
carried over the present case. According to this analysis, for a
given (sufficiently small) $b,$ there is only one stable solution.
This solution is identified as that maximizing the flame velocity.
In addition to that, the poles of the stable solution are required
to be vertically aligned in the complex $\eta$-plane. For such a
``coalescent'' solution, a simple upper bound on the number of
poles can be obtained from Eq.~(\ref{solution2}). Namely, for $k =
k_0$ with $\eta_{k_0}$ uppermost, one has
\begin{eqnarray}&&\label{solution3}
1 = \frac{\lambda_c}{2 b}\sum\limits_{\genfrac{}{}{0pt}{}{m =
1}{m\ne k_{0}}}^{2 P}\coth\left[\frac{\pi}{2 b}({\rm
Im}~\eta_{k_0} - {\rm Im }~\eta_m)\right] \ge \frac{\lambda_c}{2
b}(2 P - 1). \nonumber
\end{eqnarray}
\noindent (The equality holds, if ${\rm Im}~\eta_{k_0} = \infty$)
Then Eq.~(\ref{solution1}) tells us that the maximum $W$ is
reached for
$$P_{{\rm max}} = {\rm Int} \left(\frac{b}{\lambda_c} + \frac{1}{2}\right),$$
${\rm Int}(x)$ denoting the integer part of $x.$ Thus, the flame
velocity increase $W_s$ of the stable solution takes the form
\begin{eqnarray}&&\label{solution4}
W_s = 4 W_{{\rm max}}\frac{P_{{\rm max}}\lambda_c}{2 b}\left(1 -
\frac{P_{{\rm max}}\lambda_c}{2 b}\right),
\end{eqnarray}
\noindent where
\begin{eqnarray}\label{wmax}
W_{{\rm max}} = \frac{(\theta - 1)^2}{8\theta^2}\,.
\end{eqnarray}

\subsection{The second post-Sivashinsky approximation}\label{4order}

Before going into details of derivation of the correct
fourth-order equation, let us take a pause to identify the origin
of the failure of equations derived in Ref.~6 to satisfy the
condition (\ref{lvcondition}). It is traced to the choice of the
constant term in the decomposition Eq.~$(19)^{\prime}$ of the
potential mode$^{21}$ of the downstream flow (the formulas cited
from Ref.~6 are distinguished by the prime). In Ref.~6, this term
is taken to be equal to $\theta V.$ It reappears later in the
right hand side of Eq.~$(39)^{\prime}$ which is an integral of the
main Eq.~$(40)^{\prime}$. It was proved$^{6}$ that
Eq.~$(40)^{\prime}$ is valid up to terms of the order $\alpha^6.$
However, this does {\it not} mean that Eq.~$(39)^{\prime}$ is
valid up to terms of the order $\alpha^5.$ The point is that the
value $\theta V$ of the constant term in the potential mode is
accurate only up to $O(\alpha^3)$-terms. In other words, the
proper decomposition of $u_p$ should read
$$u_p = \theta V + \beta + \tilde{u}_p\,, \qquad \tilde{u}_p =
\int\limits_{-\infty}^{+\infty}dk\tilde{u}_k \exp(-|k|\xi + i
k\eta)\,,$$ where $\beta = O(\alpha^3)$ is some constant, instead
of Eq.~$(19)^{\prime}.$ Correspondingly, Eq.~$(39)^{\prime}$ is to
be substituted by the following
$$\phi = - \hat{H} \ln \frac{\Omega}{\theta V + \beta}\,.$$
Since $\theta, V, \Omega$ are all $O(1),$ this implies that an
additive constant of the order $O(\alpha^3)$ is missing in
Eq.~$(39)^{\prime}.$ The value of the constant $\beta$ is fixed
eventually by the condition (\ref{lvcondition}). As in the
preceding section, it can be expressed through $W$ explicitly by
averaging the resulting equation for the flame front position. It
is not difficult to verify that following this way, one obtains
exactly Eq.~(\ref{3ordereqf}) instead of Eq.~$(70)^{\prime}.$

\subsubsection{Equation for the flame front position}

Turning to the derivation of the fourth order equation, we write
\begin{eqnarray}\label{omega4}
|\omega_-|^2 = 1 + \frac{\left(f'u_-+w_-\right)^2}{N^2} = 1 +
\left(f'\right)^2 + 2f'w_- + O(\alpha^4)\,,
\end{eqnarray}
\noindent instead of Eq.~(\ref{omega3}). Substituting this into
Eq.~(\ref{main1}), using Eq.~(\ref{apb10}), and integrating yields
\begin{eqnarray}\label{main14}&&
\omega_- + \frac{\theta - 1}{2}\left(1 +
i\hat{H}\right)\left\{\frac{1-if'}{N} + \frac{\left(f'\right)^2 +
2f'w_-}{2\theta} - \frac{\lambda_c}{2\pi}f''\right\} \nonumber\\&&
- i~\frac{\theta - 1}{2\pi}\dashint\limits_{-\infty}^{+\infty}
\frac{d\bar{\eta}}{\bar{\eta} - \eta}
f''(\bar{\eta})\left[f(\bar{\eta}) - f(\eta)\right] = C\,,
\end{eqnarray}
\noindent which replaces Eq.~(\ref{main13}). Next, substituting
the third order result for $w_-$ in the curly brackets in this
equation, one finds with the required accuracy
\begin{eqnarray}&&
\omega_- - \frac{\theta - 1}{2}\left(1 +
i\hat{H}\right)\left\{\frac{if'}{N} +
\frac{\lambda_c}{2\pi}f''\right\} - i~\frac{\theta -
1}{2\pi}\dashint\limits_{-\infty}^{+\infty}
\frac{d\bar{\eta}}{\bar{\eta} - \eta}
f''(\bar{\eta})\left[f(\bar{\eta}) - f(\eta)\right] = C\,.
\nonumber
\end{eqnarray}
\noindent When extracting the real and imaginary parts, one notes
that the last term in the latter equation is purely imaginary, and
hence it affects only the $w_-$ component of the fuel velocity.
Since this term is of the fourth order, it can be omitted because
$w_-$ is multiplied by $f'$ in the evolution equation. Thus, we
have
\begin{eqnarray}&&\label{ufourth}
u_- + \frac{\theta - 1}{2}\hat{H}\left(\frac{f'}{N}\right) -
\frac{\theta - 1}{2}\frac{\lambda_c}{2\pi}f'' = C_1\,,
\end{eqnarray}
\noindent while for $w_-$ one can still use the third order
Eq.~(\ref{wthird}) (with $C_2 = 0$). Apart from the
$\varepsilon$-terms which are treated here differently from
Ref.~6, equation (\ref{ufourth}) is nothing but the expression
Eq.~$(64)^{\prime}$ for $\upsilon$ (the $\xi$-independent
counterpart of $u,$ introduced in Ref.~6), in which the constant
term is modified as discussed above. Substituting $u_-, w_-$ from
Eqs.~(\ref{wthird}), (\ref{ufourth}) into the evolution equation
(\ref{evolution1}) gives
\begin{eqnarray}&&
N  - C_1 = \frac{\theta - 1}{2}\left\{- \left(f' + \hat{H}\right)
\left(\frac{f'}{N}\right) +
\frac{\lambda_c}{2\pi}f''\right\}\,.\nonumber
\end{eqnarray}
\noindent Averaging this equation along the flame front as before,
noting that $f'(-\eta) = - f'(\eta),$ as a consequence of
Eq.~(\ref{antisym}), and applying the condition
Eq.~(\ref{lvcondition}) one can express the constant $C_1$ through
$W.$ Thus, we obtain
\begin{eqnarray}\label{4ordereqf}&& N - 1 -
\theta W = \frac{\theta - 1}{2}\left\{- \left(f' +
\hat{H}\right)\left(\frac{f'}{N}\right) +
\frac{\lambda_c}{2\pi}f''\right\}\,.
\end{eqnarray}
\noindent As in the case of the third order equations
(\ref{3ordereqf}) and $(70)^{\prime},$ the fourth order equations
(\ref{4ordereqf}) and $(73)^{\prime}$ coincide exactly upon
differentiation. To see this, one has to expand the $N$-factors in
Eq.~(\ref{4ordereqf}), to linearize the $\varepsilon$-terms in
Eq.~$(73)^{\prime},$ and to take into account that in the course
of derivation of this equation, the Sivashinsky equation was used
to transform some of the nonlinear terms.

\subsubsection{Solution of the equation (\ref{4ordereqf})}
\label{ansol4}

Remarkably, Eq.~(\ref{4ordereqf}) can be further simplified, and
brought to the form similar to that of Eq.~(\ref{3ordereqf}). This
can be done by expressing the terms of the third and fourth orders
in $f',$ which appear upon expanding the $N$-factors, through the
lower-order terms using the Sivashinsky equation. Namely,
rewriting this equation as
\begin{eqnarray}\label{2ordereqf}&&
\frac{\left(f'\right)^2}{2} = W - \frac{\theta - 1}{2}\hat{H}f' +
O(\varepsilon)\,,
\end{eqnarray}
\noindent and taking its square, the term $(f')^4/8$ coming from
the $N$ on the left of Eq.~(\ref{4ordereqf}) is substituted by
$$\frac{1}{2}\left[W^2 -
W(\theta - 1)\hat{H}f' + \frac{(\theta -
1)^2}{4}\left(\hat{H}f'\right)^2\right]\,,$$ which is already of
the second order in $f'.$ The $\varepsilon$-terms have been
omitted because we neglect all nonlinearities related to the
transport processes. Next, multiplying Eq.~(\ref{2ordereqf}) by
$f',$ and acting by the Hilbert operator, the term
$\hat{H}(f')^3/2$ appearing on the right of Eq.~(\ref{4ordereqf})
upon expanding $\hat{H}(f'/N)$ becomes
$$\hat{H}\frac{\left(f'\right)^3}{2} = W\hat{H}f'
- \frac{\theta - 1}{4}\left[\left(\hat{H}f'\right)^2 -
\left(f'\right)^2\right]\,,$$ where we have used the well-known
identity $$2\hat{H}\left(a\hat{H}a\right) =
\left(\hat{H}a\right)^2 - a^2\,.$$ After substitution of these
expressions, Eq.~(\ref{4ordereqf}) can be rewritten in the
following form, within accuracy of the fourth order,
\begin{eqnarray}\label{4ordereqff}&&
\frac{2\theta^2\left(f'\right)^2}{(\theta + 1)^2} - \left(\theta W
+ \frac{W^2}{2}\right) = \frac{\theta - 1}{2}\left( - \hat{H}f' +
\frac{\lambda_c}{2\pi}f''\right)\,.
\end{eqnarray}
\noindent

Proceeding as in Sec.~\ref{ansol}, we look for a periodic solution
of Eq.~(\ref{4ordereqff}) in the form Eq.~(\ref{anzats}), and find
\begin{eqnarray}
A &=& - \frac{\lambda_c}{2\pi}\frac{\theta - 1 }{4\theta^2}(\theta
+ 1)^2\,, \nonumber\\ \theta W + \frac{W^2}{2} &=& \frac{(\theta^2
- 1)^2}{8\theta^2 }\frac{P\lambda_c}{2 b}\left(1 -
\frac{P\lambda_c}{2 b}\right)\,, \nonumber
\end{eqnarray}
\noindent while equations for the pole positions remain the same
[see Eqs.~(\ref{solution2})]. As before, we identify the stable
solution as that maximizing the flame velocity. In particular, the
maximum of the flame velocity increase for a given $\theta$ is

$$W_{\rm max} = - \theta + \sqrt{\theta^2
+ \frac{(\theta^2 - 1)^2}{16\theta^2}}\,\,.$$ Because of the
factor $1/16,$ the second term under the square root represents a
relatively small correction to the first even for values of
$\theta$ not close to unity, so the above expression can be
simplified to
\begin{eqnarray}\label{wmax4}
W_{\rm max} = \frac{(\theta^2 - 1)^2}{32\theta^3}\,.
\end{eqnarray}
\noindent

\begin{figure}
\scalebox{0.8}{\includegraphics{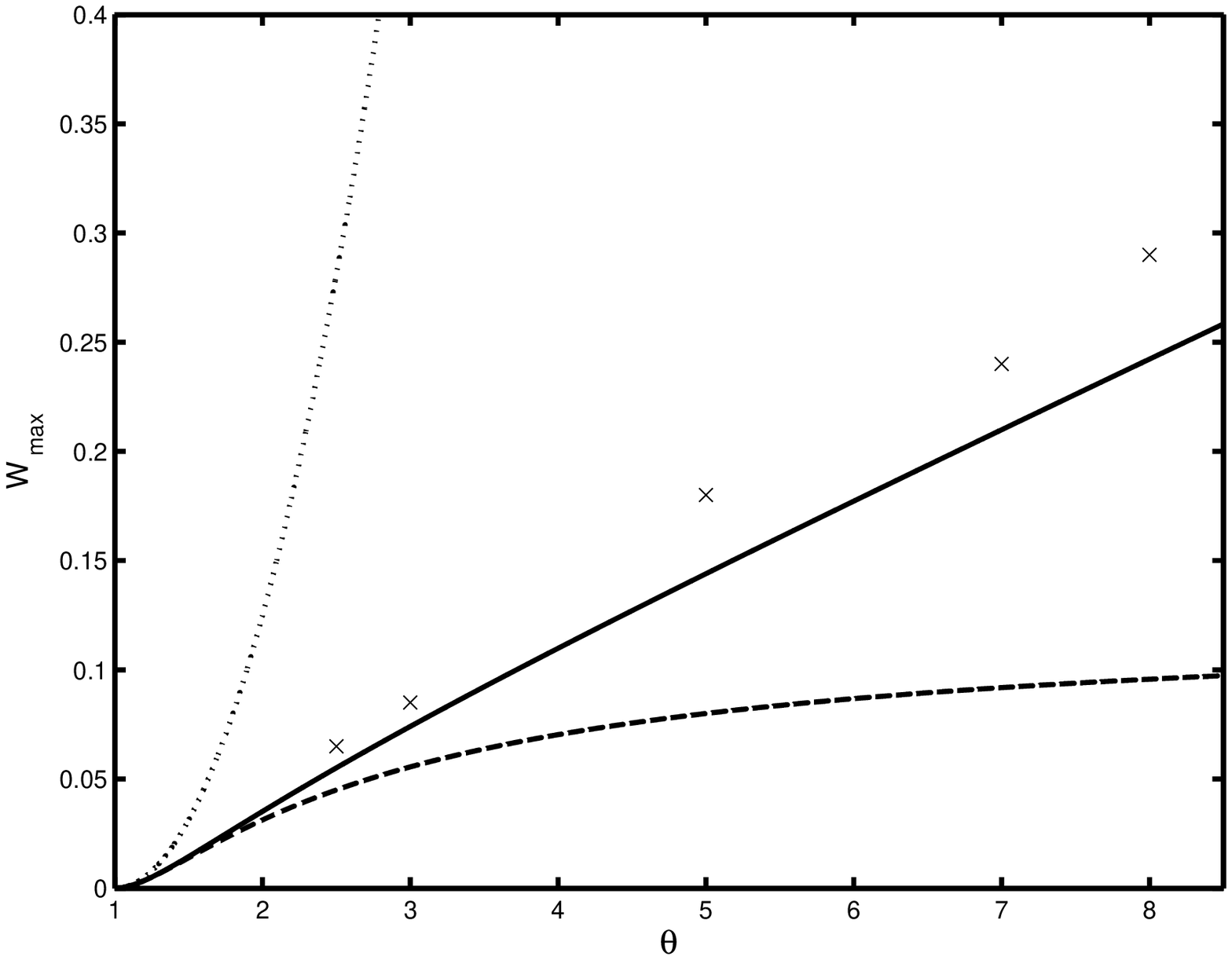}}\caption{Kazakov,
Physics of Fluids}\label{fig2}
\end{figure}

Figure~\ref{fig2} compares the dependencies of the maximal flame
velocity increase on the gas expansion coefficient, given by the
Sivashinsky equation, and its corrections, Eqs.~(\ref{3ordereqf}),
(\ref{4ordereqff}), with the experimental data$.^{9,22}$ It is
seen that the theoretical curve $W_{\rm max}(\theta)$ approaches
the experimental marks as we pass from the Sivashinsky equation to
the more accurate Eqs.~(\ref{3ordereqf}), (\ref{4ordereqff}).
However, one should remember that this improvement can be trusted
only for sufficiently small values of $(\theta - 1).$ For large
values of the gas expansion coefficient it is rather a lucky
accident. Only integration of the exact Eq.~(\ref{main1}) can give
reliable results for the practically important values of
$\theta\,.$

\section{Discussion and conclusions}\label{discussion}

The results of Sec.~\ref{closed} solve the problem of closed
description of stationary flames. The complex
Eq.~(\ref{generalc1}) determines the on-shell distribution of the
fuel velocity as a functional of the flame front configuration in
the most general form, while the evolution equation plays the role
of a consistency condition which gives an equation for the front
position itself. We have shown, furthermore, that in the case of
zero-thickness flames, the main Eq.~(\ref{generalc1}) takes the
form (\ref{main}). This equation is universal in that any surface
of discontinuity, propagating in an ideal incompressible fluid, is
described by Eq.~(\ref{main}) whatever internal structure of this
discontinuity be. The latter shows itself in the
$O(\varepsilon)$-corrections to this equation, where $\varepsilon$
is the relative thickness of the discontinuity. A simple
comparison with the results of the linear theory has shown that
the linear account of the heat conduction -- species diffusion
processes in the flame front modifies Eq.~(\ref{main}) to
Eq.~(\ref{main1}).

Next, some technical remarks are in order. It is difficult to say
at the moment whether Eq.~(\ref{main1}) admits further
simplification. What can be said, on the other hand, is that its
analytical structure does not present serious problems for
numerical integration. Indeed, the keystone of this structure is
the integral operator $\hat{\EuScript{H}}.$ But its properties are
much like those of the Hilbert operator $\hat{H}$ which is
well-known how to deal with. From the theoretical point of view,
Eq.~(\ref{main1}) is convenient for constructing various
approximate descriptions of stationary flames. In particular, it
is well suited for developing the small $(\theta - 1)$ expansion
which we have carried out in Sec.~\ref{smallt}. Specifically, it
was verified in Sec.~\ref{4order} that at the second
post-Sivashinsky approximation, the exact equations derived in the
present paper reproduce the fourth order equation obtained in
Ref.~6, up to an additive constant. This agrees completely with
the main result of Sec.~IIIB of Ref.~6, according to which
Eq.~$(40)^{\prime}$ correctly approximates the exact equation for
the flame front position up to terms of the sixth order in
$(\theta - 1).$ The difference in the values of the additive
constant was found in Sec.~\ref{4order} to be the result of an
improper integration of Eq.~$(40)^{\prime},$ namely of an
incorrect separation of a constant term in the Fourier
decomposition of the potential mode of the flow velocity
downstream.

The results presented in this paper resolve the dilemma stated in
the Introduction in the case of 2D stationary flames. Since in our
considerations we have extensively used specifically 2D
mathematical tools, the question of principle is whether the
results of this paper can be carried over to the 3D case, and
further to the general non-stationary case.

\begin{appendix}

\section{Consistency check for Eq.~(\ref{vintf})}\label{appa}

After a lengthy calculation in Sec.~\ref{structure}, we obtained
the following simple expression for the vorticity mode near the
flame front
\begin{eqnarray}\label{a1}
v^{v}_i = \int\limits_{F} d l~\chi(\varepsilon_{pq}r_p v_{q+})
\frac{v^n_+\sigma_{+}v_{i+}}{2 v^2_+}\,.
\end{eqnarray}
\noindent As this important formula plays the central role in our
investigation, a simple consistency check will be performed here,
namely, we will verify that $v^{v}_i$ given by Eq.~(\ref{a1})
satisfies
\begin{eqnarray}\label{check}
\left(\partial u^v/\partial\eta -
\partial w^v/\partial\xi\right)_+ \equiv \sigma_+\,.
\end{eqnarray}

\begin{figure}
\scalebox{0.8}{\includegraphics{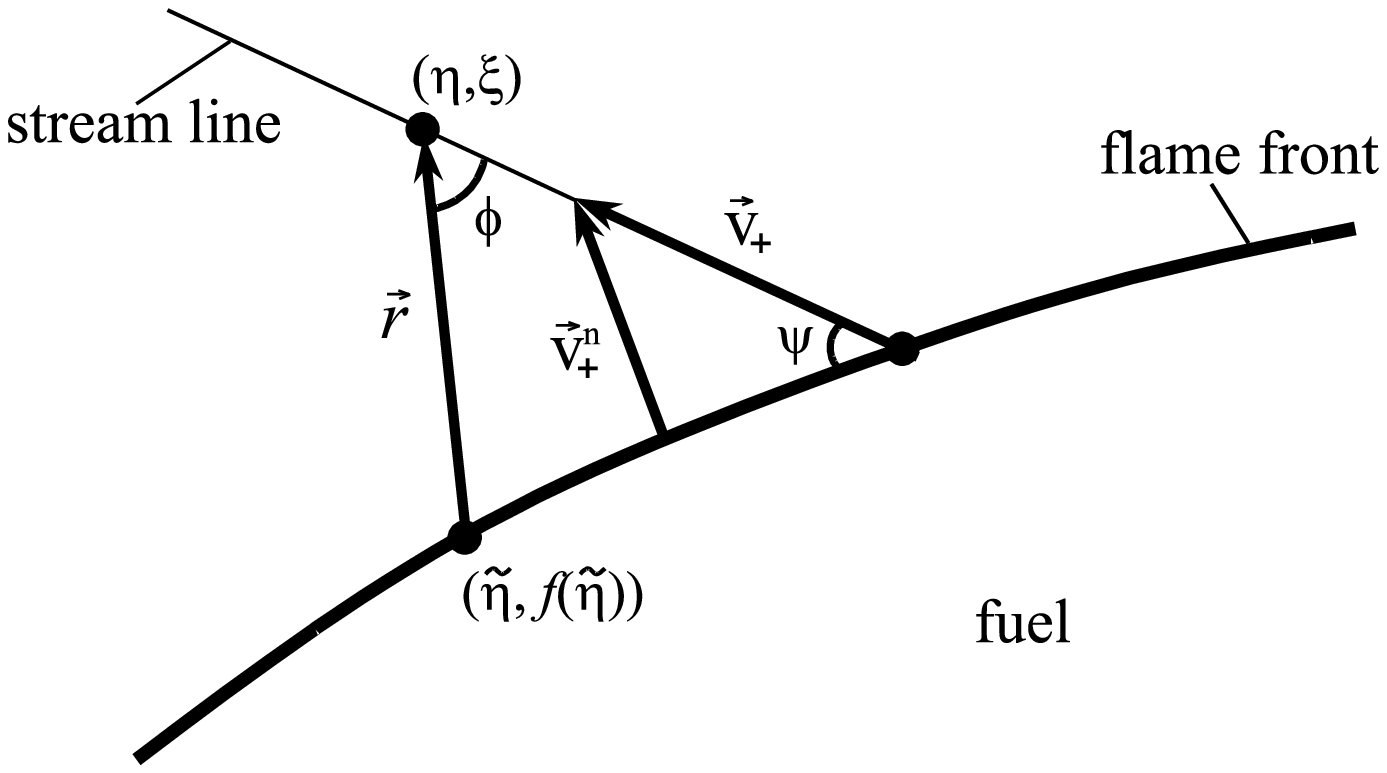}}\caption{Kazakov,
Physics of Fluids}\label{fig3}
\end{figure}

Contracting Eq.~(\ref{a1}) with $\varepsilon_{ki}\partial_k,$ and
using relation (\ref{deltad}), one finds
\begin{eqnarray}\label{a2}
\varepsilon_{ki}\partial_k v^{v}_i = \int\limits_{F} d l
~\delta(\varepsilon_{pq}r_p
v_{q+})\varepsilon_{ki}\varepsilon_{km} v_{m+}
\frac{v^n_+\sigma_{+}v_{i+}}{v^2_+} = \int\limits_{F} d l
~\delta(\varepsilon_{pq}r_p v_{q+})v^n_+\sigma_{+}\,.
\end{eqnarray}
\noindent The argument of the $\delta$-function turns into zero
when the vectors $r_i$ and $v_{i+}$ are parallel. Near this point,
one can write
$$\varepsilon_{pq}r_p v_{q+} \approx r v_{+}\phi\,,$$
where $\phi$ is the angle between the two vectors. On the other
hand, the line element, $d l,$ near the same point can be written
as
$$d l = \frac{r}{\sin\psi} d\phi = r d\phi \frac{v_{+}}{v^n_+}\,,$$
as a simple geometric consideration shows, see Fig.~\ref{fig3}.

Substituting these expressions into Eq.~(\ref{a2}), and taking
into account relation $$\delta(\alpha x) =
\frac{1}{|\alpha|}\delta(x)\,,$$ we finally arrive at the desired
identity
\begin{eqnarray}\label{a3}
\varepsilon_{ki}\partial_k v^{v}_i = \int\limits_{F} d\phi~\frac{r
v_{+}}{v^n_+} ~\delta(r v_{+}\phi)v^n_+\sigma_{+} =
\int\limits_{F} d\phi ~\delta(\phi)\sigma_{+} = \sigma_{+}\,.
\end{eqnarray}
\noindent It should be noted in this respect that the identity
Eq.~(\ref{check}) is only a necessary condition imposed on the
field $v^{v}_i.$ Playing the role of a ``boundary condition'' for
the vortex mode, Eq.~(\ref{check}) is satisfied by infinitely many
essentially different fields, {\it i.e.,} fields which are not
equal up to a potential. It is not difficult to verify, for
instance, that the velocity field defined by
\begin{eqnarray}
\tilde{v}^{v}_i = \int\limits_{F} d l~\chi(\varepsilon_{pq}r_p
n_{q}) \frac{\sigma_+ n_{i}}{2} \nonumber
\end{eqnarray}
\noindent also satisfies Eq.~(\ref{check}), and the difference
$v^{v}_i - \tilde{v}^{v}_i$ is essentially non-zero.

By the construction of Sec.~\ref{structure}, $v^{v}_i$ given by
Eq.~(\ref{a1}) is essentially the only field that satisfies the
flow equations (\ref{flow1})-(\ref{flow2}).

\section{Some properties of the operator
$\hat{\EuScript{H}}$}\label{appb}

\subsection{Proof of the identity (\ref{pr})}\label{b1}

\begin{figure}
\scalebox{0.8}{\includegraphics{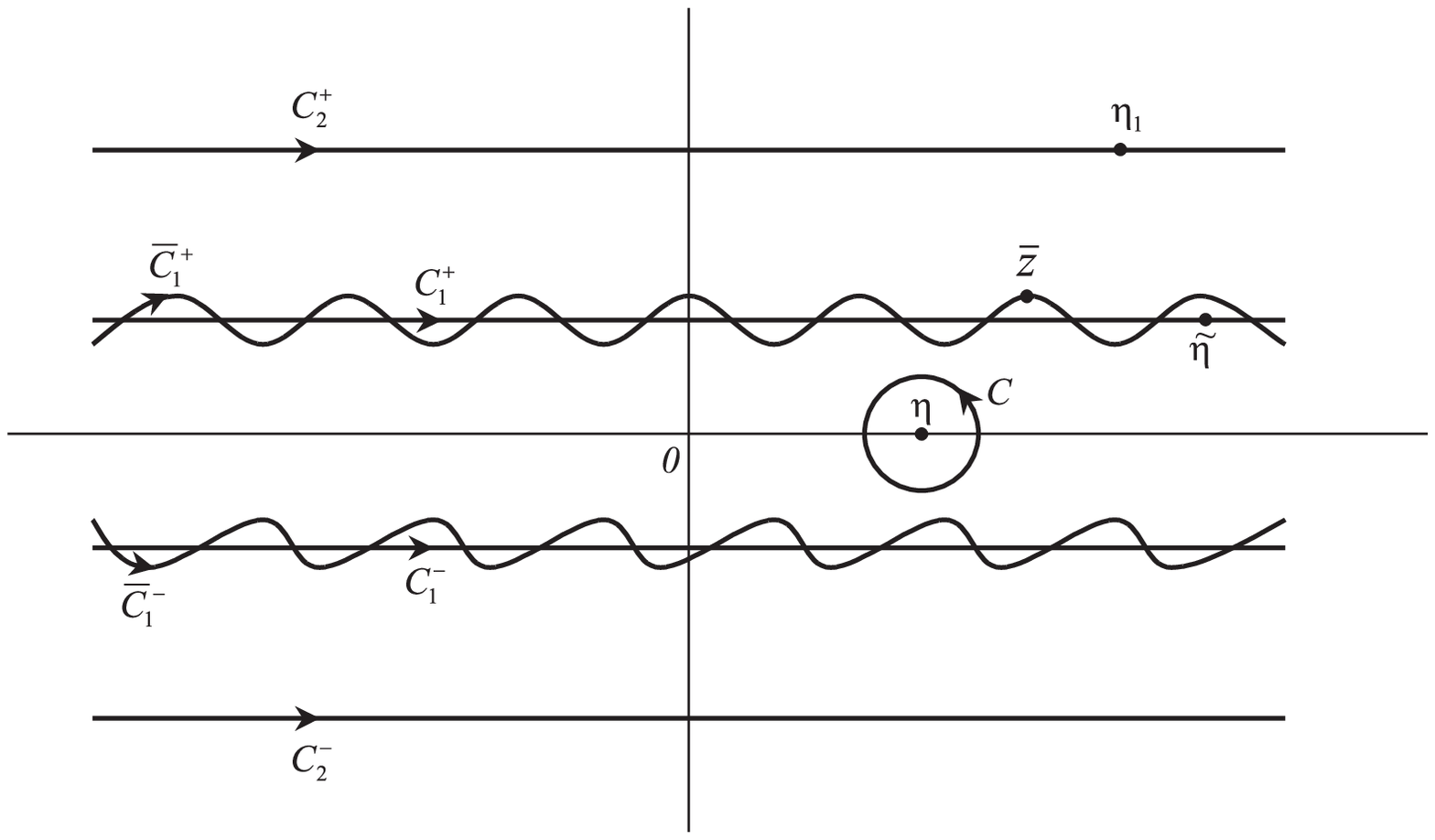}}\caption{Kazakov,
Physics of Fluids}\label{fig4}
\end{figure}

In the course of derivation of Eq.~(\ref{generalc1}), we have
introduced an operator $\hat{\EuScript{H}}$ defined on functionals
of the flow variables by
\begin{eqnarray}\label{apb1}
\left(\hat{\EuScript{H}}a\right)(\eta) = \frac{1 + i
f'(\eta)}{\pi}~\dashint\limits_{-\infty}^{+\infty}
d\tilde{\eta}~\frac{a(\tilde{\eta})}{\tilde{\eta} - \eta +
i[f(\tilde{\eta}) - f(\eta)]}\,,
\end{eqnarray}
\noindent and used repeatedly the following important identity it
satisfies $$\hat{\EuScript{H}}^2 = -1\,.$$ To prove this identity,
it is convenient to represent the right hand side of
Eq.~(\ref{apb1}) as an integral over the contour $C_1 = C_1^- \cup
C_1^+$ in the complex $\tilde{\eta}$-plane, shown in
Fig.~\ref{fig4}
\begin{eqnarray}\label{apb2}
\left(\hat{\EuScript{H}}a\right)(\eta) = \frac{1 + i
f'(\eta)}{2\pi}~\int\limits_{C_1}d
\tilde{\eta}~\frac{a(\tilde{\eta})}{\tilde{z} - z}\,,
\end{eqnarray}
\noindent where $z=\eta+if(\eta),$
$\tilde{z}=\tilde{\eta}+if(\tilde{\eta}),$ and $C_1$ is chosen
such that all singularities of the integrand, except the pole at
$\tilde{\eta} = \eta,$ remain above $C_1^+,$ or below $C_1^-.$
Under our assumption about analytical properties of the functions
involved [see discussion below Eq.~(\ref{antisym})], such contour
always exists. Now $\hat{\EuScript{H}}^2$ takes the form
\begin{eqnarray}\label{apb3}
\left(\hat{\EuScript{H}}^2 a\right)(\eta) = \frac{1 +
if'(\eta)}{4\pi^2}\int\limits_{C_1}\frac{d \tilde{z}}{\tilde{z} -
z}\int\limits_{C_2}d \eta_1~\frac{a(\eta_1)}{z_1 - \tilde{z}}\,,
\quad z_1 = \eta_1 + if(\eta_1)\,,
\end{eqnarray}
\noindent where the contour $C_2 = C_2^- \cup C_2^+$ of
integration over $\eta_1$ comprises $C_1$ (see Fig.~\ref{fig4}).
Changing the order of integration in Eq.~(\ref{apb3}), using the
formula
\begin{eqnarray}
\int\frac{d \tilde{z}}{(\tilde{z} - z)(z_1 - \tilde{z})} =
\frac{1}{z_1 - z}\ln\frac{\tilde{z} - z}{\tilde{z} - z_1}\ ,
\nonumber
\end{eqnarray}
\noindent and taking into account that the logarithm gives rise to
a nonzero contribution only if the arguments of the functions
$\tilde{z} - z$ and $\tilde{z} - z_1$ run in opposite directions
when $\tilde{z}$ runs the contours $C_1^{\pm},$ we obtain
\begin{eqnarray}&&
\frac{1 + if'(\eta)}{4\pi^2}\int\limits_{C_1}\frac{d
\tilde{z}}{\tilde{z} - z}\int\limits_{C_2}d
\eta_1~\frac{a(\eta_1)}{z_1 - \tilde{z}} = \frac{1 +
if'(\eta)}{4\pi^2}\left(2\pi i\int\limits_{C_2^-}d
\eta_1~\frac{a(\eta_1)}{z_1 - z} - 2\pi i\int\limits_{C_2^+}d
\eta_1~\frac{a(\eta_1)}{z_1 - z}\right) \nonumber\\&& = [1 +
if'(\eta)]\frac{i}{2\pi}\int\limits_C d
\eta_1~\frac{a(\eta_1)}{z_1 - z} = -[1 + if'(\eta)]\cdot{\rm
res}\left.\frac{a(\eta_1)}{(\eta_1 - \eta) + i[f(\eta_1) -
f(\eta)]}\right|_{\eta_1 = \eta} = - a(\eta)\,, \nonumber
\end{eqnarray}
\noindent and thus
$$\left(\hat{\EuScript{H}}^2 a\right)(\eta) = - a(\eta)$$
which was to be proved.

\subsection{Asymptotic form of $\hat{\EuScript{H}}$ for $\theta \to
1$}\label{b2}

It was shown in Sec.~\ref{smallt} that the exact equations derived
in Sec.~\ref{gen} constitute a general framework for perturbative
treatment of flames within the small $(\theta - 1)$-expansion. An
important step of this program is the construction of an
asymptotic expansion of the operator $\hat{\EuScript{H}}.$

Let us consider first an $\hat{\EuScript{H}}$-transform of a total
derivative (this is the case we dealt with in Sec.~\ref{smallt})
\begin{eqnarray}\label{apb4}
\left(\hat{\EuScript{H}}a'\right)(\eta) = \frac{1 + i
f'(\eta)}{\pi}~\dashint\limits_{-\infty}^{+\infty}
d\tilde{\eta}~\frac{a'(\tilde{\eta})}{\tilde{\eta} - \eta +
i[f(\tilde{\eta}) - f(\eta)]}\,.
\end{eqnarray}
\noindent The aim of the subsequent transformations will be to
develop an expansion of this integral in powers of
$[f(\tilde{\eta}) - f(\eta)]$ which is small for all
$\tilde{\eta},$ rather than in powers of $(\tilde{\eta} - \eta).$
For this purpose we first rewrite it as an integral over the
contour $C_1$ [Cf. Eq.~(\ref{apb2})]
\begin{eqnarray}\label{apb5}
\left(\hat{\EuScript{H}}a'\right)(\eta) = \frac{1 + i
f'(\eta)}{2\pi}~\int\limits_{C_1}\frac{d
a(\tilde{\eta})}{\tilde{z} - z}\,.
\end{eqnarray}
\noindent Integrating by parts then yields
\begin{eqnarray}\label{apb6}
\left(\hat{\EuScript{H}}a'\right)(\eta) =
\frac{1}{2\pi}\frac{d}{d\eta}~\int\limits_{\tilde{C}_1}
d\tilde{z}~\frac{a(\tilde{\eta}(\tilde{z}))}{\tilde{z} - z}\,,
\end{eqnarray}
\noindent where $\tilde{C}_1$ denotes a contour in the complex
$\tilde{z}$-plane, which is run by $\tilde{z} = \tilde{\eta} + i
f(\tilde{\eta})$ when $\tilde{\eta}$ runs $C_1.$ Next, let us make
the following change of the integration variable in
Eq.~(\ref{apb6})
$$\tilde{z} \to \bar{z} = \tilde{z} - i f(\eta)\,.$$ This gives
\begin{eqnarray}\label{apb7}
\left(\hat{\EuScript{H}}a'\right)(\eta) =
\frac{1}{2\pi}\frac{d}{d\eta}~\int\limits_{\bar{C}_1}
d\bar{z}~\frac{a(\tilde{\eta}(\bar{z}))}{\bar{z} - \eta}\,,
\end{eqnarray}
\noindent where the contour $\bar{C}_1$ of $\bar{z}$-integration
is obtained by a uniform vertical shift of $\tilde{C}_1$ by an
amount $f(\eta).$ Note that $\bar{z} = \tilde{\eta} + i
[f(\tilde{\eta}) - f(\eta)] = \tilde{\eta} + O(\alpha).$ Hence,
within the framework of the asymptotic expansion in $\alpha,$
$\bar{C}_1$ is to be considered as a small deformation of the
initial contour $C_1.$ Taking $\alpha$ small enough we can always
secure $C_1$ from crossing singularities of the integrand
(including the pole $\tilde{\eta} = \eta$) under this deformation.
Assuming this, we deform $\bar{C}_1$ back to the real axis, and
obtain
\begin{eqnarray}\label{apb8}
\left(\hat{\EuScript{H}}a'\right)(\eta) =
\frac{1}{\pi}\frac{d}{d\eta}~\dashint\limits_{-\infty}^{+\infty}
d\bar{\eta}~\frac{a(\tilde{\eta}(\bar{\eta}))}{\bar{\eta} -
\eta}\,,
\end{eqnarray}
\noindent where $\tilde{\eta}(\bar{\eta})$ is a function of the
real variable $\bar{\eta},$ defined by
$$\bar{\eta} = \tilde{\eta} + i [f(\tilde{\eta}) - f(\eta)].$$
Resolving this relation with respect to $\tilde{\eta}$ in the form
of a series
\begin{eqnarray}\label{apb9}
\tilde{\eta} &=& \bar{\eta} - i [f(\tilde{\eta}) - f(\eta)]
\nonumber\\ &=& \bar{\eta} - i [f(\bar{\eta}) -
f(\eta)] - f'(\bar{\eta})[f(\tilde{\eta}) - f(\eta)]
+ \frac{i}{2} f''(\bar{\eta})[f(\tilde{\eta}) - f(\eta)]^2 + \dots \nonumber\\
&=& \bar{\eta} - i [f(\bar{\eta}) - f(\eta)] -
f'(\bar{\eta})[f(\bar{\eta}) - f(\eta)] + i
\left[f'(\bar{\eta})\right]^2 [f(\bar{\eta}) - f(\eta)]
\nonumber\\&& + \frac{i}{2}f''(\bar{\eta})[f(\bar{\eta}) -
f(\eta)]^2 + \dots\,,
\end{eqnarray}
\noindent and substituting it into Eq.~(\ref{apb8}), we finally
arrive at the following expansion of $\hat{\EuScript{H}}a'$
\begin{eqnarray}\label{apb10}
\left(\hat{\EuScript{H}}a'\right)(\eta) &=&
\frac{1}{\pi}\frac{d}{d\eta}~\dashint\limits_{-\infty}^{+\infty}
\frac{d\bar{\eta}}{\bar{\eta} - \eta}~\Bigl\{a(\bar{\eta}) - i
a'(\bar{\eta})[f(\bar{\eta}) - f(\eta)] -
a'(\bar{\eta})f'(\bar{\eta})[f(\bar{\eta}) - f(\eta)]
\nonumber\\&& + i a'(\bar{\eta})\left[f'(\bar{\eta})\right]^2
[f(\bar{\eta}) - f(\eta)] -
\frac{1}{2}a''(\bar{\eta})[f(\bar{\eta}) - f(\eta)]^2 +
\dots\Bigr\}\,.
\end{eqnarray}
\noindent Since each term in the curly brackets in the integrand
on the right of Eq.~(\ref{apb10}) is periodic, the corresponding
improper integrals are well-defined by the rule (\ref{rulec}).

Now it is straightforward to write down the result of asymptotic
action of $\hat{\EuScript{H}}$ on a general $a(\eta).$ Namely,
using $d/d\eta\hat{H} = \hat{H}d/d\eta$ in the first term in
Eq.~(\ref{apb10}), and making the substitution $a'(\eta)\to
a(\eta)$ throughout this equation yields
\begin{eqnarray}\label{apb11}
\left(\hat{\EuScript{H}}a\right)(\eta) &=&
\left(\hat{H}a\right)(\eta) +
\frac{1}{\pi}\frac{d}{d\eta}~\dashint\limits_{-\infty}^{+\infty}
\frac{d\bar{\eta}}{\bar{\eta} - \eta}~\Bigl\{- i
a(\bar{\eta})[f(\bar{\eta}) - f(\eta)] -
a(\bar{\eta})f'(\bar{\eta})[f(\bar{\eta}) - f(\eta)] \nonumber\\&&
+ i a(\bar{\eta})\left[f'(\bar{\eta})\right]^2 [f(\bar{\eta}) -
f(\eta)] - \frac{1}{2}a'(\bar{\eta})[f(\bar{\eta}) - f(\eta)]^2 +
\dots\Bigr\}\,.
\end{eqnarray}
\noindent

In the course of derivation of the third order equation in
Sec.~\ref{smallt}, we found it necessary to determine the
asymptotic action of $\hat{\EuScript{H}}$ on the curly bracket in
the left hand side of Eq.~(\ref{main1}) taking into account
$O(\alpha^3)$ terms. In this case, $$a' = - if'' + u_-' =
O(\alpha^2)\,,$$ and therefore, it is sufficient to keep only the
first term in the integrand of Eq.~(\ref{apb10}), which gives
immediately
\begin{eqnarray}\label{apb12}
\left(\hat{\EuScript{H}}a'\right)(\eta) =
\left(\hat{H}a'\right)(\eta) + O(\alpha^4)\,.
\end{eqnarray}

\end{appendix}

$^1$G.~I.~Sivashinsky, ``Nonlinear analysis of hydrodynamic
instability in laminar flames,'' Acta Astronaut. {\bf 4}, 1177
(1977).

$^2$M.~Matalon and B.~J.~Matkowsky, ``Flames as gasdynamic
discontinuities,'' J.~Fluid Mech. {\bf 124}, 239 (1982).

$^3$P.~Pelce and P.~Clavin, ``Influences of hydrodynamics and
diffusion upon the stability limits of laminar premixed flames,''
J.~Fluid Mech. {\bf 124}, 219 (1982).

$^4$G.~I.~Sivashinsky and P.~Clavin, ``On the nonlinear theory of
hydrodynamic instability in flames,'' J.~Physique {\bf 48}, 193
(1987).

$^5$K.~A.~Kazakov and M.~A.~Liberman, ``Effect of vorticity
production on the structure and velocity of curved flames,''
Phys.~Rev.~Lett. {\bf 88}, 064502 (2002).

$^6$K.~A.~Kazakov and M.~A.~Liberman, ``Nonlinear equation for
curved stationary flames,'' Phys.~Fluids {\bf 14}, 1166 (2002).

$^7$L.~D.~Landau, ``On the theory of slow combustion,'' Acta
Physicochimica URSS {\bf 19}, 77 (1944).

$^8$G.~Darrieus, unpublished work presented at La Technique
Moderne, and at Le Congr${\rm\grave e}$s de M${\rm\acute
e}$canique Appliqu${\rm\acute e}$e, (1938) and (1945).

$^9$V.~V.~Bychkov, S.~M.~Golberg, M.~A.~Liberman, and
L.~E.~Eriksson, ``Propagation of curved stationary flames in
tubes,'' Phys.~Rev.~E {\bf 54}, 3713 (1996).

$^{10}$We do consider zero-thickness flames in Sec.~\ref{zero},
but the only purpose of this consideration is to simplify
treatment of the non-zero case.

$^{11}$Ya.~B.~Zel'dovich, A.~G.~Istratov, N.~I.~Kidin, and
V.~B.~Librovich, ``Flame propagation in tubes: hydrodynamics and
stability,'' Combust.~Sci.~and~Tech. {\bf 24}, 1 (1980).

$^{12}$Indeed, the second term is a pure gradient, while the curl
of the first term is proportional to $\Delta\ln r$ which is equal
to zero everywhere in the bulk, Cf. Eq.~(\ref{green}).

$^{13}$As to dispersion relation for the gas velocity upstream, it
can be written directly in terms of $\omega$ itself, since the
analog of Eq.~(\ref{cauchy}) in this case is free of any
divergences {\it a priori}, and gives in the limit $z\to F,$
$$\omega(z_-) = - \frac{1}{\pi i}\dashint\limits_{F}
\frac{d\tilde{z}}{\tilde{z} - z_-}\omega(\tilde{z}) + V\,.$$

$^{14}$Ya.~B.~Zel'dovich, G.~I.~Barenblatt, V.~B.~Librovich, and
G.~M.~Makhviladze, {\it The Mathematical theory of combustion and
explosion} (Consultants Bureau, New-York, 1985) pp. 466-470.

$^{15}$Using this equation one should remember that the evolution
equation still has the form (\ref{evolution1}) [all
$\varepsilon$-contributions are already collected in the right
hand side of Eq.~(\ref{main1})].

$^{16}$G.~Joulin and P.~Cambray, ``On a tentative, approximate
evolution equation for markedly wrinkled premixed flames,''
Combust.~Sci.~and~Tech. {\bf 81}, 243 (1992).

$^{17}$Another way to show that $C_2 = 0$ is to recall that no
terms of the form $C_2 f'$ appear in the linear equation for the
flame front position. However, we prefer the reasoning given in
the text, because it avoids referring back to the linear theory.

$^{18}$O.~Thual, U.~Frish, and M.~Henon, ``Application of pole
decomposition to an equation governing the dynamics of wrinkled
flames,'' J.~Phys. (France) {\bf 46}, 1485 (1985).

$^{19}$G.~Joulin, ``On the Zhdanov-Trubnikov equation for premixed
flame stability,'' J. Exp. Theor. Phys. {\bf 73}, 234 (1991).

$^{20}$D.~Vaynblat and M.~Matalon, ``Stability of pole solutions
for planar propagating flames: I. Exact eigenvalues and
eigenfunctions \& II. Properties of eigenvalues and eigenfunctions
with implication to flame stability,''  SIAM J. Applied Math. {\bf
60}, 679, 703 (2000).

$^{21}$The same problem with the Sivashinsky-Clavin equation$^{4}$
was shown$^{16}$ to be of a similar origin.

$^{22}$M.~A.~Liberman {\it et al.,} ``Numerical studies of curved
stationary flames in wide tubes,'' Combust. Theory Modelling {\bf
7}, 653 (2003).

\newpage

\centerline{\large Figure captions}

Fig.1: Elementary decomposition of the flow downstream.

Fig.2: Dependence of the maximal flame velocity increase on the
gas expansion coefficient, given by the Sivashinsky equation
(dotted line), and its corrections -- Eq.~(\ref{3ordereqf})
(dashed line), and Eq.~(\ref{4ordereqff}) (full line). The marks
represent the results of numerical experiment$^{9,22}$ (accuracy
of the experimental data is about $10\%$).

Fig.3: Near-the-front structure of the flow downstream.
$\bm{v}^n_+ = \bm{n} v^n_+$ is the normal component of the
velocity. Since the observation point $(\eta,\xi)$ is close to the
flame front, the stream line and the part of the front near this
point can be considered straight.

Fig.4: Contours of integration in Eqs.~(\ref{apb2}) --
(\ref{apb7})

\clearpage

\clearpage

\clearpage

\clearpage

\end{document}